\documentclass[fleqn,usenatbib]{mnras}

\usepackage{newtxtext,newtxmath}

\usepackage[T1]{fontenc}
\usepackage{ae,aecompl}

\usepackage{graphicx}	
\usepackage{amsmath}	
\usepackage{amssymb}	
\usepackage{float}
\usepackage[caption = false]{subfig}
\usepackage{pdflscape}
\title[RCW~103 with Chandra and XMM-Newton]{
Progenitors and Explosion Properties of Supernova Remnants Hosting Central Compact Objects: I. RCW~103 Associated with the Peculiar Source 1E~161348--5055}

\author[C. Braun et al.]{
C. Braun,$^{1}$\thanks{E-mail: umbrau59@myumanitoba.ca}
S. Safi-Harb,$^{1}$\thanks{E-mail: samar.safi-harb@umanitoba.ca}
C. L. Fryer$^{2,3,4,5}$\thanks{E-mail: fryer@lanl.gov} 
\\
$^{1}$Department of Physics and Astronomy, University of Manitoba, Winnipeg, MB R3T 2N2, Canada \\
$^2$ Center for Theoretical Astrophysics, Los Alamos National Laboratory, Los Alamos,
NM 87545\\
$^3$ Department of Astronomy, The University of Arizona, Tucson,
AZ 85721\\
$^4$ Department of Physics and Astronomy, The University of
New Mexico, Albuquerque, NM 87131\\
$^5$ Department of Physics, The George Washington University, Washington, DC 20052
}

\date{To appear in MNRAS}

\pubyear{2018}

\begin{document}
\label{firstpage}
\pagerange{\pageref{firstpage}--\pageref{lastpage}}
\maketitle

\begin{abstract}
We present a \textit{Chandra} and \textit{XMM-Newton} imaging and spectroscopic study of the supernova remnant (SNR) RCW~103 (G332.4--00.4) containing the Central Compact Object 1E~161348--5055. The high resolution \textit{Chandra} X-ray images reveal
enhanced emission in the south-eastern and north-western regions. Equivalent width line images of Fe~L, Mg, Si, and S using \textit{XMM-Newton} data were used to map the distribution of ejecta. The SNR was sectioned into 56 regions best characterized by two-component thermal models. The harder component ($\text{kT}\sim0.6$~keV) is adequately fitted by the VPSHOCK non-equilibrium ionization model with an ionization timescale n$_\text{e}\text{t}\sim10^{11}$--$10^{12}$~cm$^{-3}$~s, and slightly enhanced abundances over solar values. The soft component ($\text{kT}\sim0.2$~keV), fitted by the APEC model, is well described by plasma in collisional ionization equilibrium with abundances consistent with solar values. Assuming a distance of 3.1~kpc and a Sedov phase of expansion into a uniform medium, we estimate an SNR age of 4.4~kyr, a swept-up mass M$_{\text{sw}}=16$~$f_s^{-1/2}$~D$_{3.1}^{5/2}$~M$_\odot$, and a low explosion energy E$_*=3.7\times10^{49}$~$f_s^{-1/2}$~D$_{3.1}^{5/2}$~erg. This energy could be an order of magnitude higher if we relax the Sedov assumption, the plasma has a low filling factor, the plasma temperature is under-estimated, or if the SNR is expanding into the progenitor's wind-blown bubble. Standard explosion models did not match the ejecta yields. By comparing the fitted abundances to the most recent core-collapse nucleosynthesis models, our best estimate yields a low-mass progenitor around 12--13~M$_\odot$, lower than previously reported. We discuss degeneracies in the model fitting, particularly the effect of altering the explosion energy on the progenitor mass estimate.
\end{abstract}




\section{Introduction}

    X-ray observations of supernova remnants (SNRs) represent one of the most important means to study the intrinsic properties of supernova explosions, the distribution of ejecta, the nature of their collapsed cores, and the conditions of the interstellar medium (ISM). The morphology and dynamics of SNRs are highly shaped by both the progenitor star and the ISM. Explosions of massive stars result in ejecta and a shock wave that propagate through the ISM, creating a shell-like structure that emits radiation detectable in the X-ray band. The nucleosynthesis products of supernovae have prominent emission lines in the 0.3--10~keV range, which are referred to as the oxygen-group and intermediate mass elements	(O, Ne, Mg, Si, S, Ar, Ca) and the iron-group elements (mostly Fe and Ni). Furthermore, the X-ray emission lines provide information about the temperature and ionization state of the hot plasma which can be used to infer the supernova explosion properties.  
	
    RCW~103 is a young, Galactic shell-type SNR with a hard X-ray point source close to its centre, labelled 1E~161348--5055 (hereafter 1E1613) \citep{Tuohy}. A kinematic distance of 3.3~kpc was determined by \cite{Carter} from the 21 cm hydrogen-line absorption along the line of sight of the remnant. However, \cite{Leibowitz} found a visual extinction of 4.5~mag and derived a value of 6.6~kpc. The most recent distance calculation using the HI 21cm line determined a distance of 3.1~kpc by \cite{Reynoso}, and it is this value that will be used throughout the paper. 
	
    Radio studies indicate a nearly circular, 10$^{\prime}$ shell characterized by a spectral index of $\alpha\sim0.5$ (S$_{\nu}\sim\nu^{-\alpha}$) and a flux density of 28~Jy at 1~GHz \citep{Green}. Radio studies show a thick shell of a fairly young age that has only recently transitioned from the double-shock phase of its evolution \citep{Dickel}. The most recent radio study by \cite{Paron} report on the detection of HCO$^+$ and \textsuperscript{12}CO emission in the vicinity of the southern shock front providing evidence of an interaction between the SNR and a molecular cloud. Much like in radio, infrared (IR) observations of the H$_2$ 2.122~$\mu$m line and other IR lines also found evidence of molecular cloud interactions \citep{Burton, Oliva1990, Oliva1999, Rho}. A more recent IR study from \textit{Spitzer} shows more diffuse emission from the 24~$\mu$m line similar to the limb-brightened X-ray morphology \citep{Reach, Pinheiro}. An OH~(1720 MHz) maser line detection also shows evidence of a molecular cloud interaction in the south \citep{Frail}. Optical filaments were detected in the X-ray limbs, most prominently in the southern limb \citep{Van, Ruiz}. An optical study comparing photographic plates estimated an expansion rate of 1100~km~s$^{-1}$ with an age estimate of 2000~yr \citep{Carter}.
	
    It's worth noting that the SNR has a `bilateral' morphology, with an axis of symmetry running from north-east to south-west. A recent study by \cite{2016A&A...587A.148W} examined all Galactic SNRs at radio wavelengths and compiled a list of SNRs with a `bilateral' morphology, of which RCW~103 is one of them. The study concluded that the morphology of the majority of the `bilateral' SNRs sample is influenced by the Galactic magnetic field at the SNR location, and was also shown to be the case for RCW~103. 

    Past X-ray studies focused on the unusual compact object, 1E1613, at the remnant's centre. This compact object, classified as a Central Compact Object (CCO),  displays strong X-ray variability \citep{Gotthelf} and a periodicity of 6.67~hr \citep{DeLuca2006} The long $\sim$6.7~hr periodicity	suggested several possibilities: an accreting binary system, such as a young pre-low-mass X-ray binary system with an eccentric orbit operating in the `propeller' phase \citep{Bhadkamkar, Reynoso, DeLuca2007}, a magnetar born in a low-mass binary system \citep{Pizzolato}, or an isolated highly magnetized neutron star accreting material from a fossil disk resulting from fallback after the neutron star's birth and slowing it down to an extremely slow period \citep{Li, Ikhsanov, DeLuca2007}. Recently, 1E1613 went into a period of bursting activity detected by Swift/BAT and followed up by telescopes in X-rays (\textit{Chandra}, \textit{NuSTAR} and \textit{Swift}) and in the optical/NIR \citep{Rea, 2016MNRAS.463.2394D, Tendulkar}. The multi-wavelength properties of the bursting source were found to be consistent with those of a magnetar, and so the accreting binary scenario was ruled out and the 6.67-hr periodicity has been interpreted as the rotation period of the slowest known magnetar to date. The uniqueness of this CCO is our primary motivation to perform a dedicated study of its hosting SNR, RCW~103, especially as the X-ray emission from the remnant hasn't received much attention. 

    An early \textit{Einstein} observation of the SNR RCW~103 \citep{Nugent} was fit with a single non-equilibrium ionization (NEI) plasma model, and the X-ray spectrum was shown to have roughly solar abundance values of Mg, Si, S and Fe with plasma temperature $\text{kT} \sim 0.51$~keV. The X-ray emission was interpreted as consistent with shocked circumstellar medium with approximately solar composition of heavy elements. A \textit{Chandra} study of 24 SNRs including RCW~103 \citep{Lopez} found enhanced abundances for Mg, Si, and Fe; however the scope of this study focused on morphological classification and typing of SNRs rather than a dedicated spectroscopic study. Finally, a study using \textit{Chandra} data was more recently done by \cite{Frank} and will be contrasted to this study in detail in $\S$5. 		
	
    In this paper, archived \textit{Chandra} and \textit{XMM-Newton} data are used to perform an X-ray imaging and spectroscopic study of the remnant to accompany the extensive work done on its CCO. The primary motivation for our study is to address the properties of the SN explosion and progenitor that created this unique CCO. Our goals of the data analysis are to: (1) perform a spatially resolved spectroscopic study of the SNR to determine the plasma temperatures, ionization timescales, chemical abundances and their distribution across the remnant, (2) infer the supernova explosion properties (explosion energy and progenitor mass) by comparing the ejecta abundances to the newest nucleosynthesis model yields, and (3) constrain the ambient density and SNR age, as well as verify its distance estimate.

    The paper is organized as follows. In $\S$2 we summarize the observations. In $\S$3 and \S4, we describe the \textit{Chandra} and \textit{XMM-Newton} imaging and spectroscopic study, respectively.	Finally, in $\S$5 we discuss our results and in $\S$6 we summarize our conclusions.
	
    This work is part of a more global study of SNRs, to be presented in follow-up studies, using X-ray imaging and spectroscopy aimed at addressing the SN progenitors and explosion properties of SNRs hosting a diversity of compact objects, including CCOs \citep{2017JPhCS.932a2005S}. Our study is additionally motivated by performing a systematic study using the latest nucleosynthesis models available to the SNR community, while also providing feedback to modellers given the current limitations and assumptions made on nucleosynthesis model yields.

\section{Observations and Data Reduction}

	\begin{table*}
		\begin{center}
		
		\caption{Observation Data }
		\label{tbl:ExpTime}
    		\begin{tabular}{c c c c c } 
				\hline
			
				ObsID    & 
				Detector & 
				DATAMODE &
				\begin{tabular}[c]{@{}c@{}} Effective   \\ Exp. Time \\ (ks) \end{tabular} & \begin{tabular}[c]{@{}c@{}} Observation \\ Date \\ (DD/MM/YY) \end{tabular} \\
		
				\hline
				
				123   & ACIS--I & FAINT & 13.36 & 26/06/99 \\
				970   & ACIS--S & FAINT & 17.46 & 08/08/00 \\
				11823 & ACIS--I & FAINT & 62.47 & 01/06/10 \\
				12224 & ACIS--I & FAINT & 17.82 & 27/06/10 \\
				17460 & ACIS--I & VFAINT & 24.76 & 13/01/15 \\  
 				0113050601 & MOS~1/2 & Full Frame & 16.0/15.2 & 03/09/01 \\
				0113050701 & MOS~1/2 & Full Frame & 12.4/9.4 & 03/09/01 \\
				0302390101 & MOS~1/2 & Full Frame & 60.2/55.0 & 23/08/05 \\
				\hline
			\end{tabular}
		\end{center}
	\end{table*}

    \subsection{\textit{Chandra}}

    The SNR RCW~103 was observed with the \textit{Chandra} X-ray Observatory on five separate occasions as listed in Table~\ref{tbl:ExpTime}. The telescope has 2 sets of imaging CCDs as part of the Advanced CCD Imaging Spectrometer (ACIS) detector which has an energy range between 0.3--10~keV and an energy resolution of 130~eV at 1.49~keV for the front-illuminated chips and an energy resolution of 95~eV at 1.49~keV for the back-illuminated chips. Four of the observations were performed using ACIS-I which utilizes the four front-illuminated CCDs arranged in a two by two grid. The SNR spans all four chips, leaving gaps in the image data where the chip edges meet. ObsID~970 uses a single back-illuminated CCD with the ACIS-S3 chip, resulting in some of the SNR's edges falling off the chip and leaving small regions of the SNR undetected. Regions for study were carefully selected to avoid chip gaps across the different data sets (see \S4).

    Data reduction and analysis were performed using the \textit{Chandra} Interactive Analysis of Observations (CIAO) Version 4.9 software. The data were reprocessed in accordance with the \textit{CIAO} data preparation thread to reprocess the level 2 X-ray data. Periods of high background rates were removed and the effective exposure times are given in Table~\ref{tbl:ExpTime}. Using the CIAO command \textit{wavdetect}, external sources were detected and then removed. When performing the spectroscopic study in $\S$4, we selected and subtracted source-free background regions for each region. These background regions were carefully selected such that they were close to the SNR and the region of interest, as well as falling on the same chip as the selected region. The background regions are the same across all data, with an exception for ObsID 970. ObsID 970 backgrounds were chosen as source-free background regions selected from the same chip that the SNR is located (the ACIS-S3 chip). Multiple background regions were also examined, yielding similar results as expected. The spectral analysis was performed using XSPEC version 12.9.1 with the spectra binned using a minimum of 20 counts per bin. The data sets were modelled separately and then simultaneously fit for 54 regions and for the entire SNR. 

    \subsection{\textit{XMM-Newton}}

	\textit{XMM-Newton} observed RCW~103 on three separate occasions, ObsID 0113050601 (601), ObsID 0113050701 (701), and obsID 030239010 (101). We used the European Photon Imaging Camera (EPIC), which has two Metal Oxide Semi-conductor (MOS) CCD cameras covering the energy range between 0.2 and 12~keV with an energy resolution of 0.15~keV at 1~keV \citep{EPIC_MOS}. The CCDs of the pn camera were in Small Window mode and did not cover the entire SNR and so were not used in this study. The data were reprocessed using the \textit{XMM-Newton} Science Analysis System (SAS) version 16.0.0 and the most recent calibration files. The event files were reprocessed using the SAS task \textit{emchain}. The MOS data were filtered to retain patterns 0--12, to the energy range of 200--12000 eV, using the \textit{\#\! XMMEA\textunderscore EM} flag, and finally to remove any flaring yielding the effective exposure times shown in Table~\ref{tbl:ExpTime}. Given \textit{XMM-Newton's}
	superior sensitivity to low-surface brightness regions, and \textit{Chandra} being superior to \textit{XMM-Newton} for the spatially resolved spectroscopic analysis, we restrict our \textit{XMM-Newton} analysis to the equivalent width maps and to spectroscopy of selected low-surface brightness regions that would complement our \textit{Chandra} analysis.
	
    We created continuum-subtracted line images for Mg, Si, S and the Fe-L blend, as well as the equivalent width images (EWI) with their respective energy ranges, as summarized in Table~\ref{tbl:LineData} and presented in $\S$3. The data were also used to extract spectra from selected low-surface brightness regions in the SNR (see \S4.2 and Fig.~7). External sources were detected using the \textit{edetect\textunderscore chain} command and filtered out of the image using \textit{evselect}. Finally, spectra were extracted as described by the SAS handbook, with no pile-up detected. Redistribution matrix files (RMF) and ancillary response files (ARF) were created using the \textit{rmfgen} and \textit{arfgen} commands respectively, with "extendedsource=yes" flagged for the \textit{arfgen} command. The spectra were then grouped and binned to a minimum of 20 counts per bin using the FTOOL \textit{grppha}. The spectral analysis is presented in $\S$4.

\section {X-ray Imaging}
	
	\subsection{\textit{Chandra}}
	
\begin{figure*}
		\begin{center}
			\includegraphics[width=.7\textwidth]{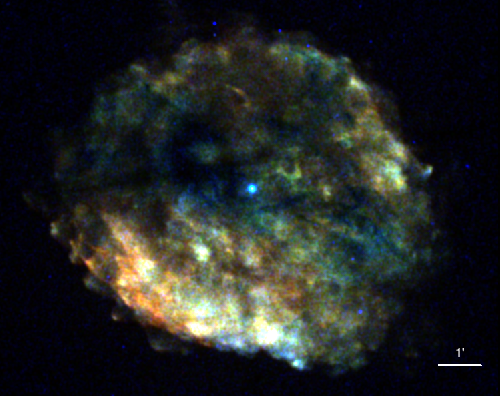}	
			\caption{RGB \textit{CHANDRA} image of RCW~103, with red, green and blue colours corresponding to the energy ranges 0.5--1.2~keV, 1.2--2.0~keV, and 2.0--7.0~keV, respectively. The image has been smoothed using a Gaussian with a radius of 3 pixels. North is up and east is to the left. An obvious artifact is the ``crosshairs'' shape almost through the centre that is due to the chip gaps from one of the data sets. }
			\label{fig:RGB}
		\end{center}
	\end{figure*}

	In X-rays, the remnant has a nearly circular morphology with a diameter of 10$^{\prime}$ and an inhomogeneous interior. Fig.~\ref{fig:RGB} shows the \textit{Chandra} RGB image with red, green and blue corresponding to the the soft (0.5--1.2~keV), medium (1.2--2.0~keV) and hard (2.0--7.0~keV) energy bands, respectively. This image reveals small-scale, clumpy structures throughout as well as two large, brighter regions in the south-east and north-west. The southern limb has some soft sections in the east and harder sections moving into the west, with multi-band (white) knots throughout. The northern limb also has some multi-band clumps but appears harder and fainter than the southern limb. North-east of the CCO, a `C-shaped' hole of X-ray emission is seen (in both the RGB and broadband images), with overall much fainter emission in the north-eastern section of the SNR. Finally, the CCO located towards the SNR centre stands out in the hard X-ray band as the blue point-like source.

\begin{figure*}
			\newpage
			\begin{center}
				\includegraphics[width=.7\textwidth]{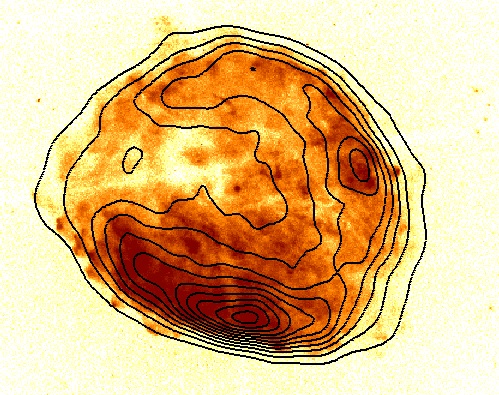}	\caption{ \textit{CHANDRA} X-ray broadband (0.3--10~keV) image  overlaid with radio contours from the Molonglo Observatory Synthesis Telescope. 10 contours are shown in logarithmic scale ranging in levels from 0.05 to 1.7 counts per pixel.}
				\label{fig:Radio}
			\end{center}
	\end{figure*}

	In Fig.~\ref{fig:Radio} we show the \textit{CHANDRA} broadband (0.3--10~keV) image with the Molonglo Observatory Synthesis Telescope (\textit{MOST}) radio contours at 843~MHz overlaid. The radio emission mimics the same overall shape as the X-ray image with contours following the large, bright southern limb and the smaller, north-western limb. While the peculiar `C-shape' X-ray hole evident in the X-ray images is as not clearly visible in the radio data, the radio emission is weaker in that region. The outermost radio shell, which likely indicates the location of the forward shock, overlaps with the outermost diffuse emission in X-rays. As for other CCOs, the CCO 1E1613 is completely absent at radio wavelengths. 

    \subsection{\textit{XMM-Newton}}

   The EWIs were constructed using the method developed by \cite{HwangCasA} and are commonly used for identifying ejecta over a wide range of surface brightnesses \citep{Frank, Park, Hwang}. The data were first filtered as described in $\S$2, then combined using the SAS command \textit{merge}. Images were constructed using the \textit{evselect} command for a pixel size of 2$^{\prime\prime}$ (40 counts per bin), and restricted to the energy ranges as described in Table~\ref{tbl:LineData} such that each line had a narrow continuum band on either side of the specified line, called the ``shoulders''. The shoulders were used to linearly interpolate the continuum over the energy range of the line, pixel by pixel, using the FTOOL \textit{farith}. The constructed continuum images and the line images were smoothed using a Gaussian with radius of 3 pixels and then the continuum was subtracted from the line images to create the continuum-subtracted line images (setting any negative counts to 0) as shown in Fig.~\ref{fig:EWI}. To further eliminate any continuum contributions from the line images, we divide the continuum-subtracted line images by the continuum images to construct the EWIs. To avoid enhancing low continuum regions in the EWIs, the EWI pixel value was set to 0 if the continuum value was below 15\% of the mean counts from the continuum image for each line. The EWI values were also set to 0 if they were negative. The final images were smoothed with a Gaussian of radius 3 pixels for Mg and Fe, and a Gaussian of radius 5 pixels for Si and S and can also be found in Fig.~\ref{fig:EWI}.

    The line images show that the distribution of the elements is well correlated with the broadband image, with the majority of emission in the lobes and with particularly bright emission in the south-eastern lobe. The EWIs do not follow this lobed structure and show a more or less uniform morphology. The Fe L image shows a flat distribution across the remnant, with no obvious bright knots of emission and with a relative depression in the north-eastern part. Mg seems to be anti-correlated from the lobed structure, with almost no emission in the southern lobe, however the bright knot near the CCO remains, as well as enhanced emission in the low-surface brightness regions in the north-east and south-west. Si, to a lesser extent than Mg, is also anti-correlated to the lobe structure, but has a bright knotty structure throughout. S has poorer statistics than the other lines but appears to have some brighter emission to the south-west.

	\begin{table}
		\begin{center}
		   	\caption{Line emission and the low and high continuum energy bands selected to construct the line images and the equivalent width images. }
		   	\label{tbl:LineData}
			\begin{tabular}{c c c c } 
				
				\hline
				
				Atomic Line  & 
			    \begin{tabular}[c]{@{}c@{}} Line \\ (keV) \end{tabular}&
				\begin{tabular}[c]{@{}c@{}} Low \\ (keV) \end{tabular} & \begin{tabular}[c]{@{}c@{}} High \\ (keV) \end{tabular} \\
		
				\hline
				
				Fe L & 0.81--0.97 & 0.75--0.80 & 1.10--1.16 \\
				Mg   & 1.28--1.43 & 1.21--1.26 & 1.43--1.48 \\
				Si   & 1.76--1.95 & 1.68--1.75 & 1.96--2.05 \\
				S    & 2.37--2.56 & 2.27--2.36 & 2.57--2.68 \\
				\hline
			\end{tabular}
		\end{center}
	\end{table}
	
	\begin{figure*}
		\begin{center}
		\subfloat[Fe L]{\includegraphics[width=0.39\textwidth]{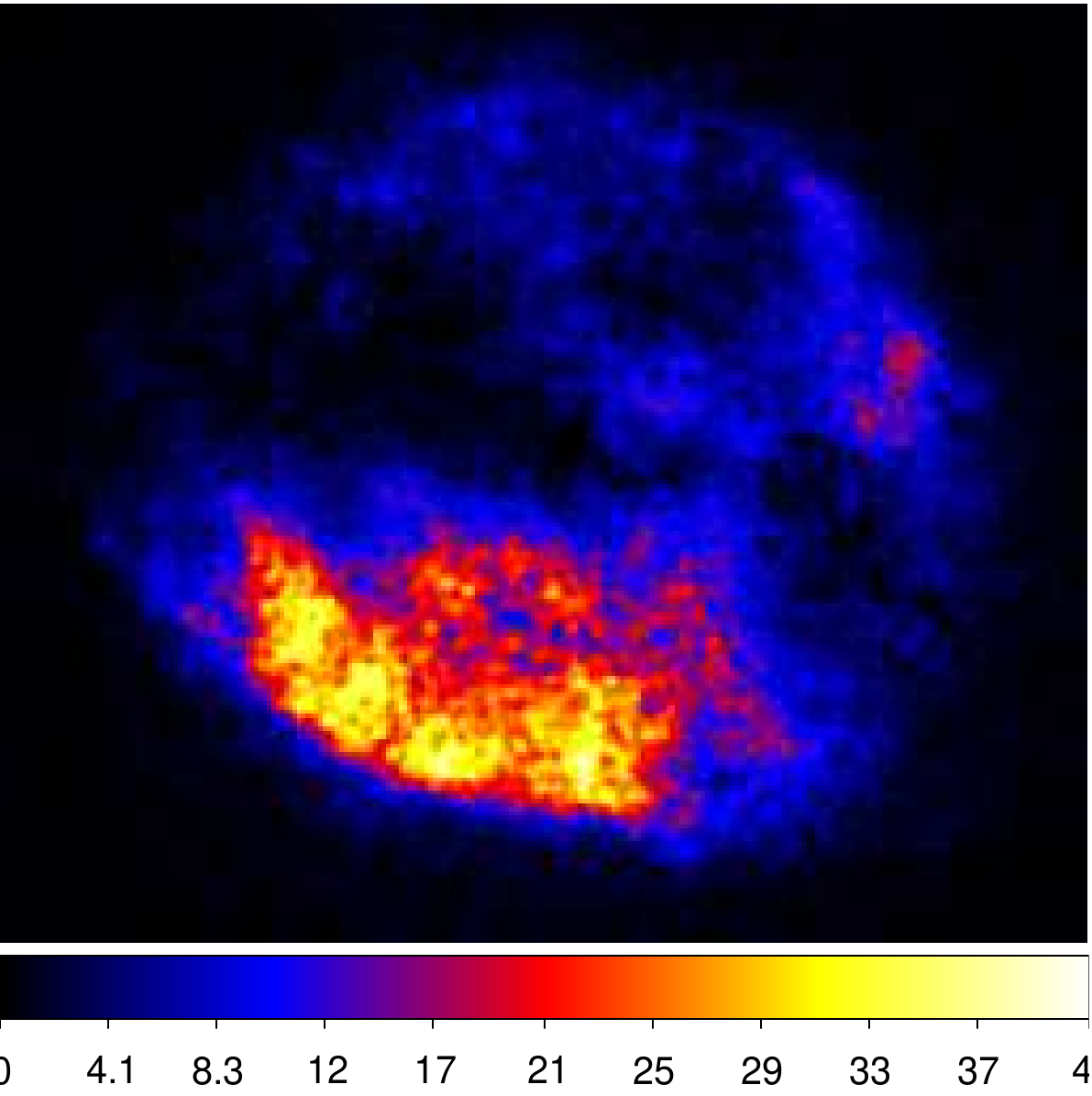}}
		\subfloat[Mg]{\includegraphics[width=0.39\textwidth]{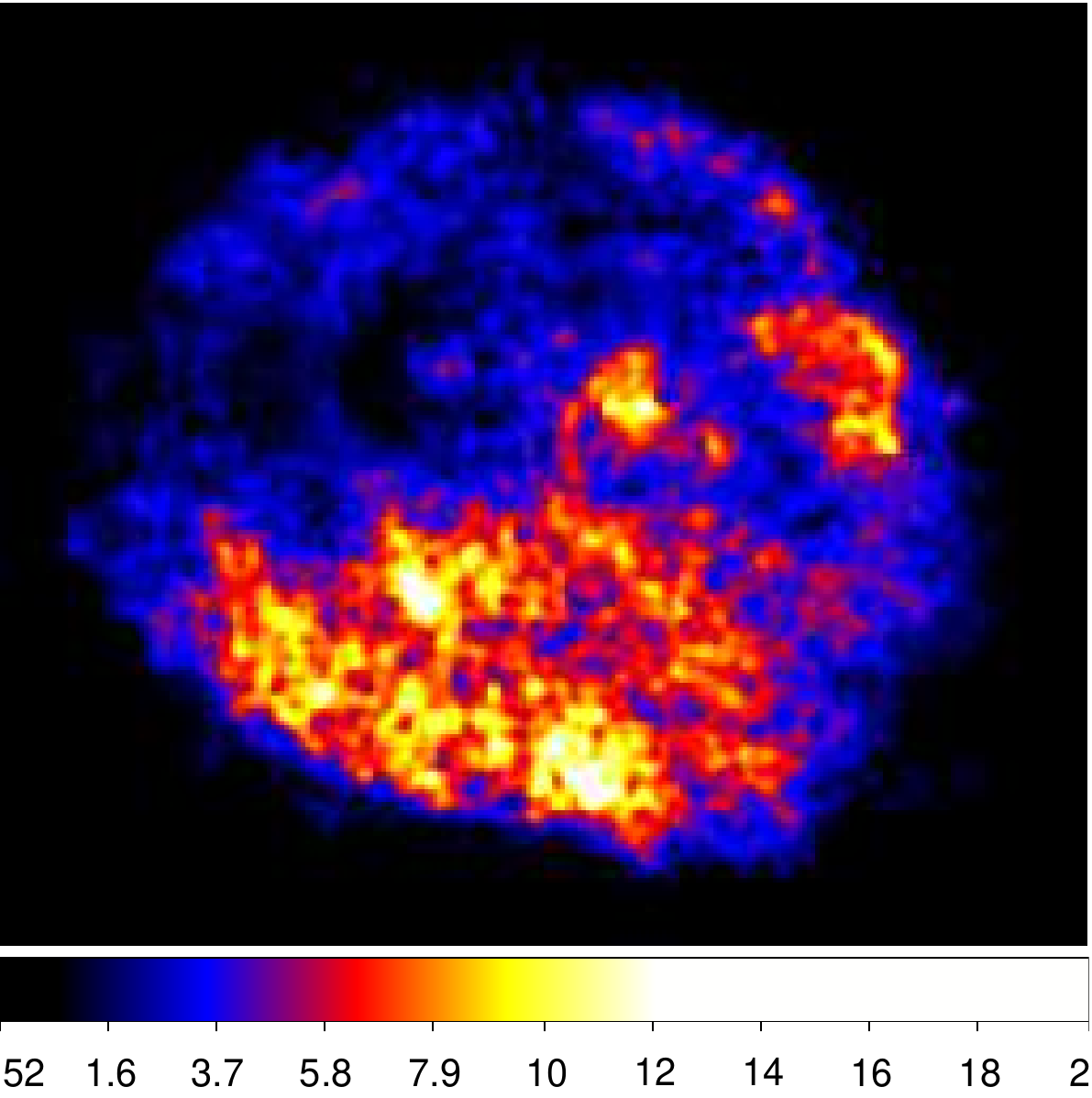}}
		\\
		\subfloat[Si]{\includegraphics[width=0.39\textwidth]{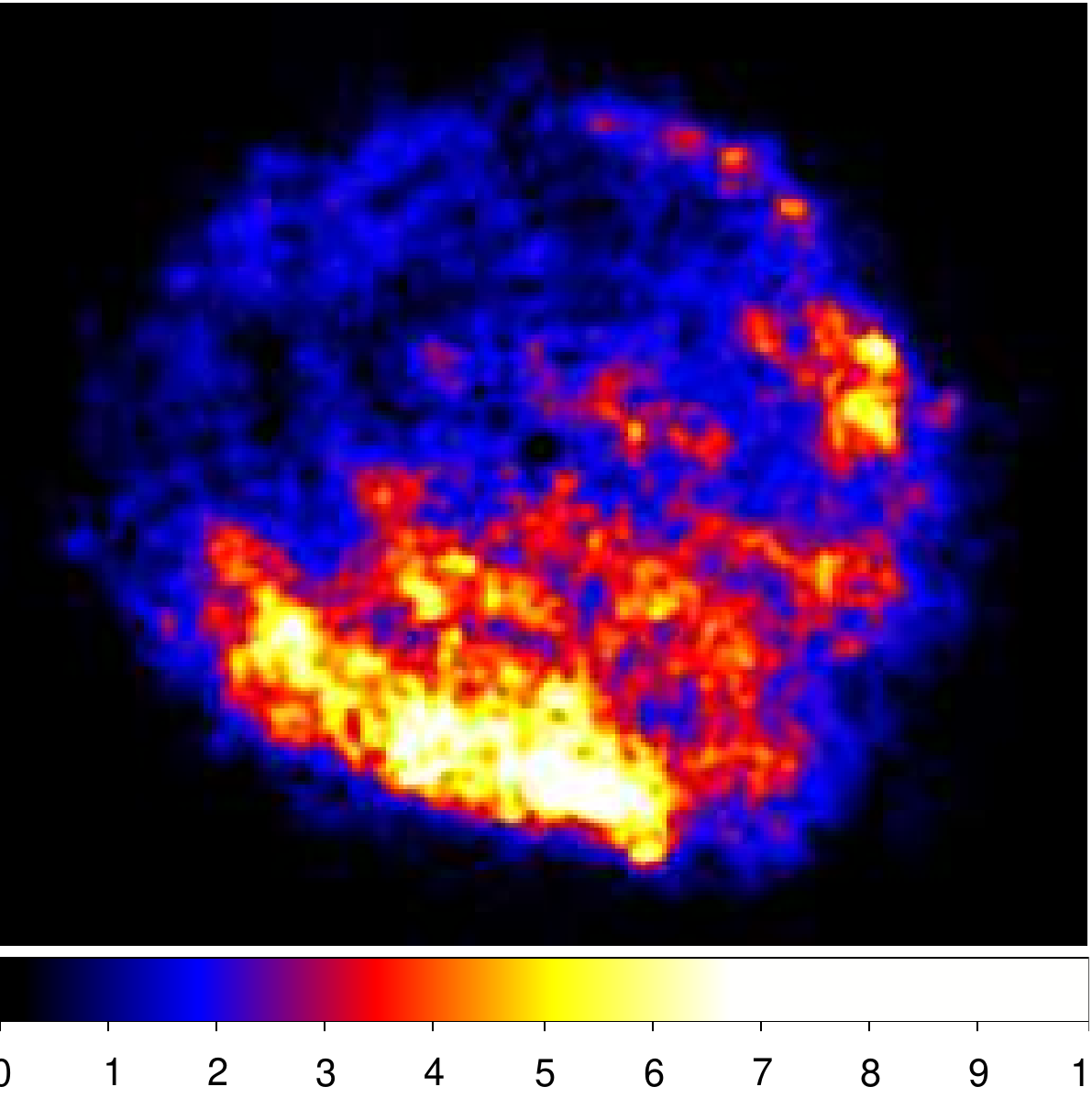}}
		\subfloat[S]{\includegraphics[width=0.39\textwidth]{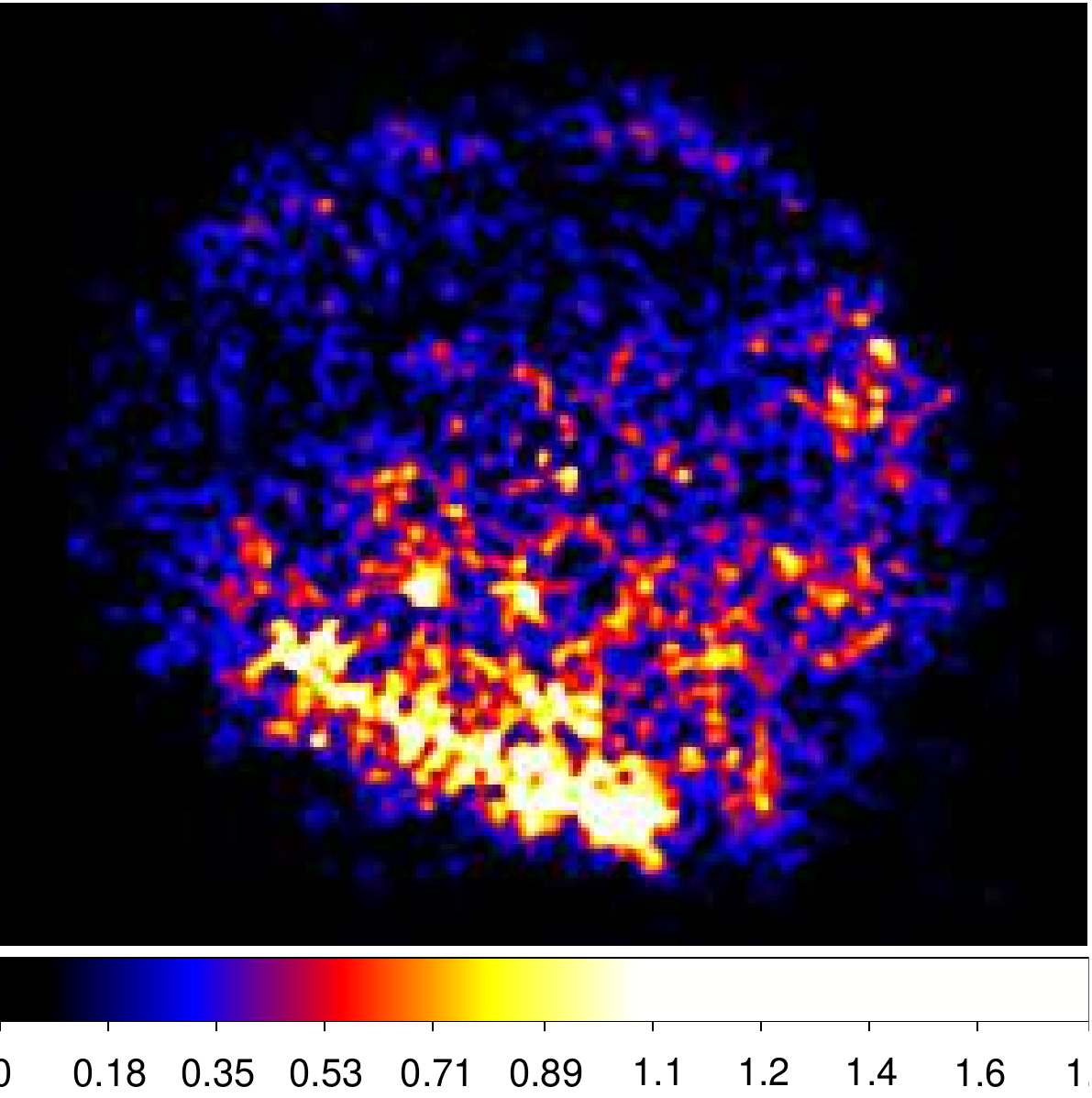}}
		\\
		\subfloat[Fe L EWI]{\includegraphics[width=0.39\textwidth]{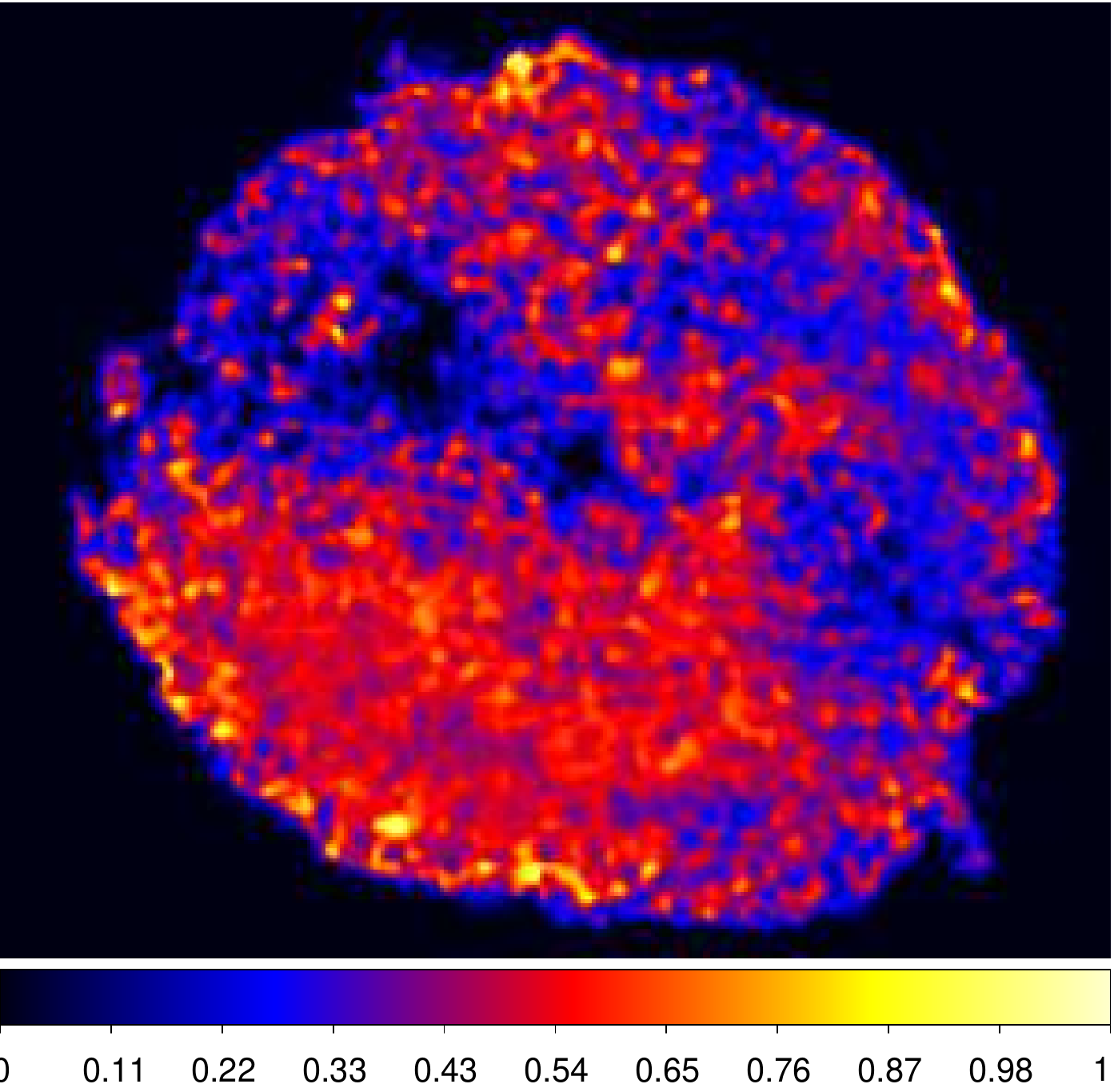}}
		\subfloat[Mg EWI]{\includegraphics[width=0.39\textwidth]{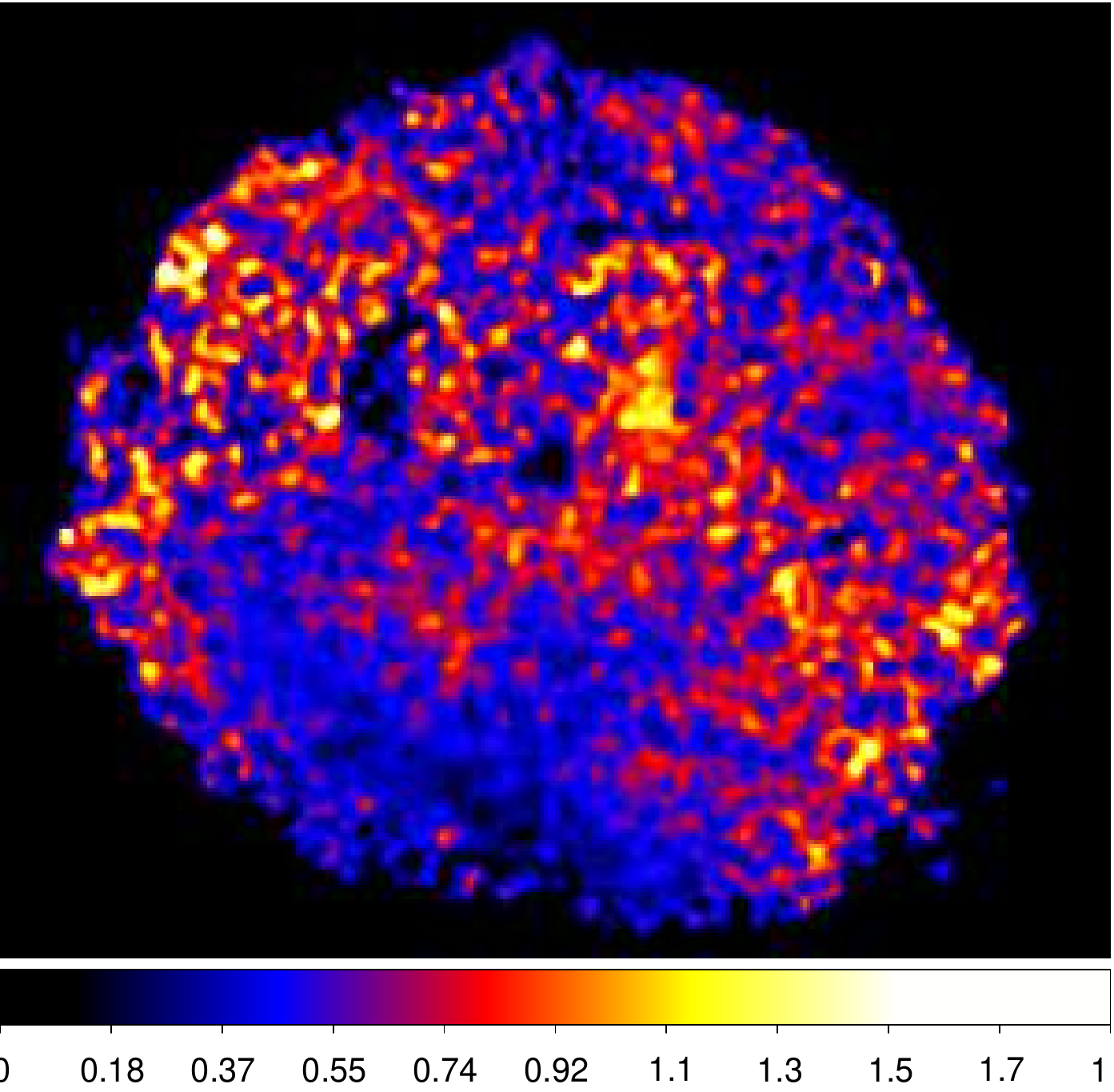}} 
		\end{center}
    \end{figure*}
	\begin{figure*}
	    \begin{center}
		\subfloat[Si EWI]{\includegraphics[width=0.39\textwidth]{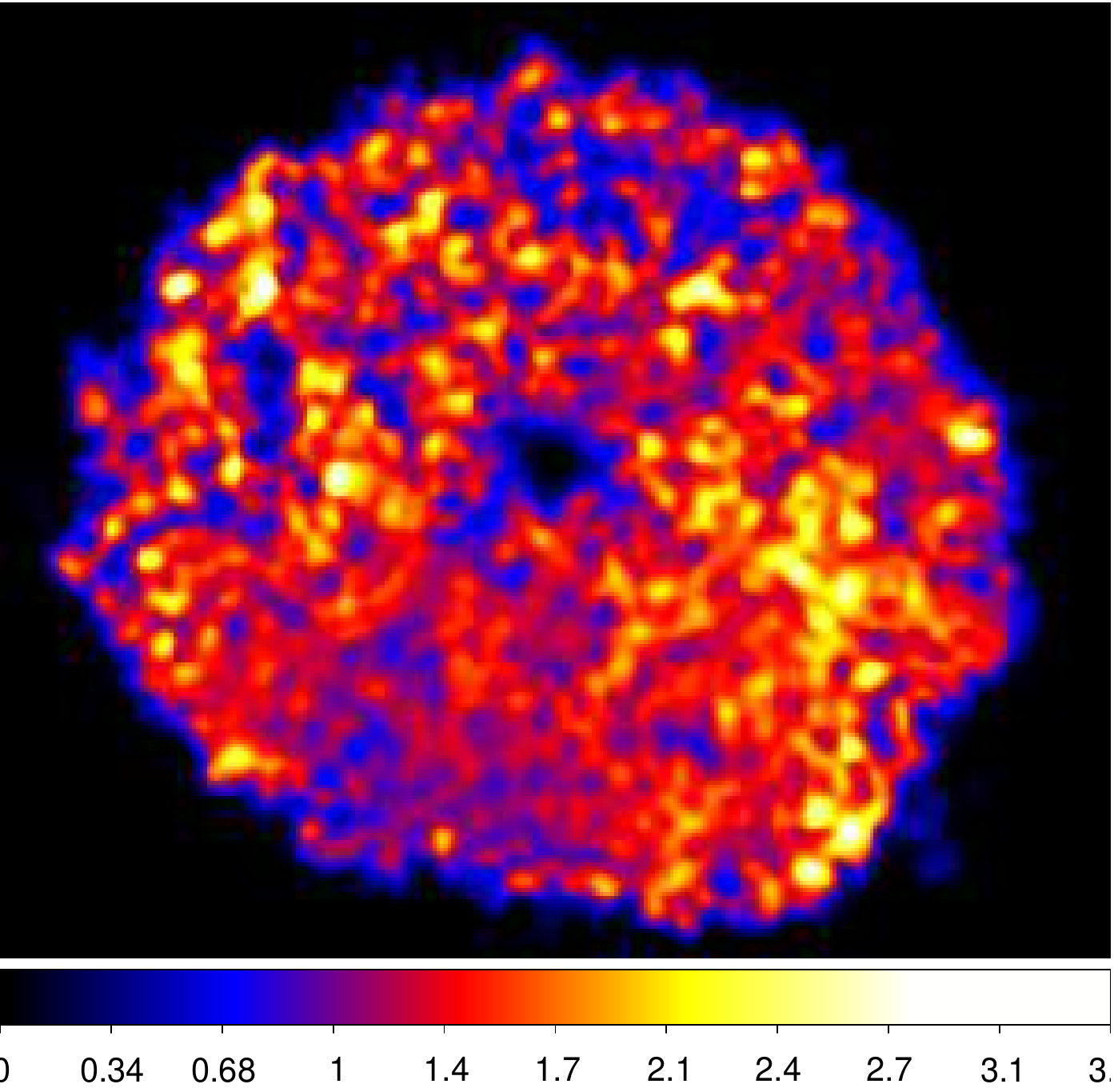}}
		\subfloat[S EWI]{\includegraphics[width=0.39\textwidth]{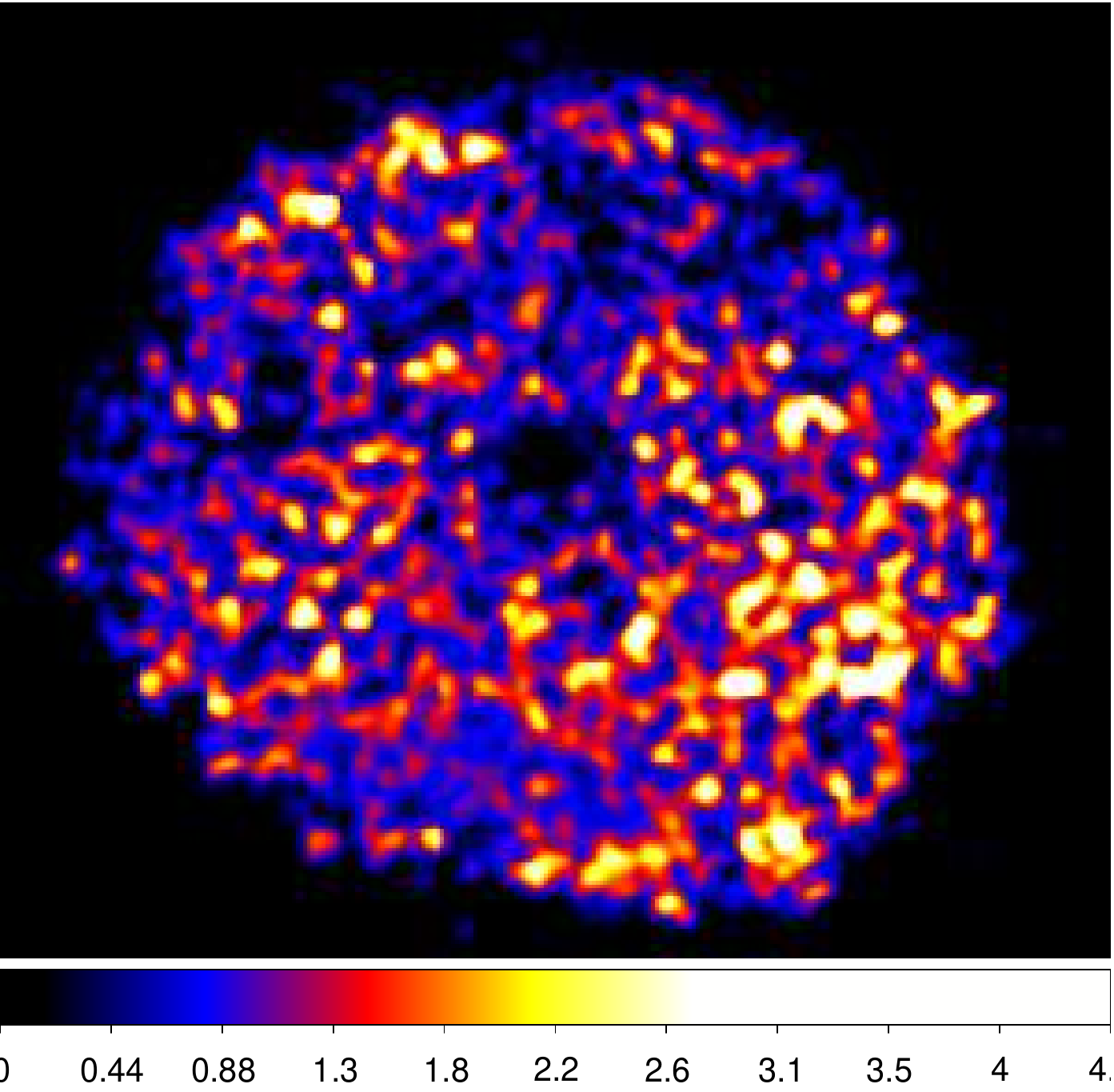}}
        \caption{Continuum-subtracted line images (top row) and equivalent width images (EWI, bottom row) for Fe~L, Mg, Si, and S. The EWIs for Fe L and Mg are smoothed with a Gaussian of radius 3 pixels whereas the Si and S EWIs are smoothed with a Gaussian of radius 5 pixels. }
		\label{fig:EWI}
		\end{center}
	\end{figure*}

	\begin{figure*}
		\begin{center}
			\includegraphics[width=0.75\textwidth]{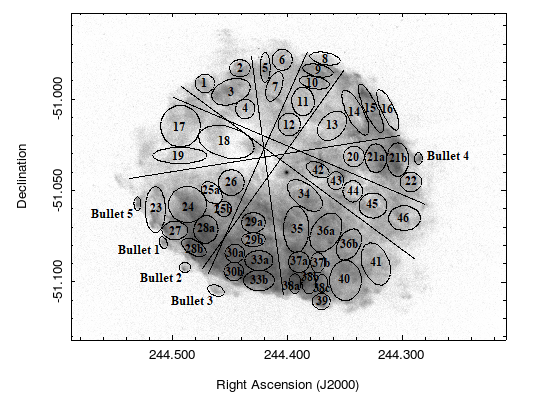}	
			\caption{The regions selected for the spatially resolved spectroscopic study,
			covering the whole SNR. 
			Lines across the image represent the gaps between the chips, and the studied regions were selected to avoid these gaps. }
			\label{fig:Regions}
		\end{center}
	\end{figure*}	
		
\section{Spectral Analysis}
	
	\subsection{\textit{Chandra}}
	
	In the following section, we perform a spatially resolved spectroscopic analysis of the SNR combining all the \textit{Chandra} data, with the exception of the data set ObsID 123, which was excluded due to contaminants. The regions were selected to cover the entire SNR, while avoiding the chip gaps related to each observation. These regions can be found in Fig.~\ref{fig:Regions}. Spectra extraction for each of the regions was completed using the \textit{CIAO} task command \textit{specextract}. The spectrum for RCW~103 is dominated by thermal emission, with emission from the Fe L blend, Mg, Si and S lines. The spectral data were restricted to between 0.5--5.0~keV. Energies higher than 5.0~keV were omitted from the spectral fits due to poor signal-to-noise ratio while at energies lower than 0.3~keV, the molecular build up on the optical filters is the most severe. Data were fit using the X-ray spectral fitting software XSPEC. The abundance tables were set using the XSPEC command \textit{abund wilm} \citep{Wilms}. During the fitting process, a multiplicative constant was introduced to account for the differences between the data sets. The ObsIDs' 12224 and 11823 constants were set to 1, where the constant for the obsID 970 was allowed to vary due to the difference in chip types, and the constant for ObsID~17460 was allowed to vary due to the different datamode. Another multiplicative model TBABS is used to account for any X-ray absorption and is characterized by the molecular hydrogen column density, N$_\text{H}$. We tested two different plasma models: a non-equilibrium ionization	model, VPSHOCK, appropriate for modelling young SNRs whose plasma has not yet reached ionization equilibrium, and	the APEC model, an emission spectrum from collisionally-ionized diffuse gas calculated from the AtomDB atomic database. VPSHOCK is a plane-parallel plasma model with variable abundances characterized by a constant electron temperature, kT, and a range of ionization timescales, n$_\text{e}$t, where n$_\text{e}$ is the post-shock electron density and t is the time since the passage of the shock. The APEC model
	is characterized by a constant electron temperature, kT, a single abundance variable, and with no ionization timescale associated with it since plasma has reached collisional ionization equilibrium (CIE).

\begin{figure*}
		\begin{center}
		\subfloat[(a) Region 16 (VPSHOCK+APEC)]{\includegraphics[angle=0,width=0.38\textwidth]{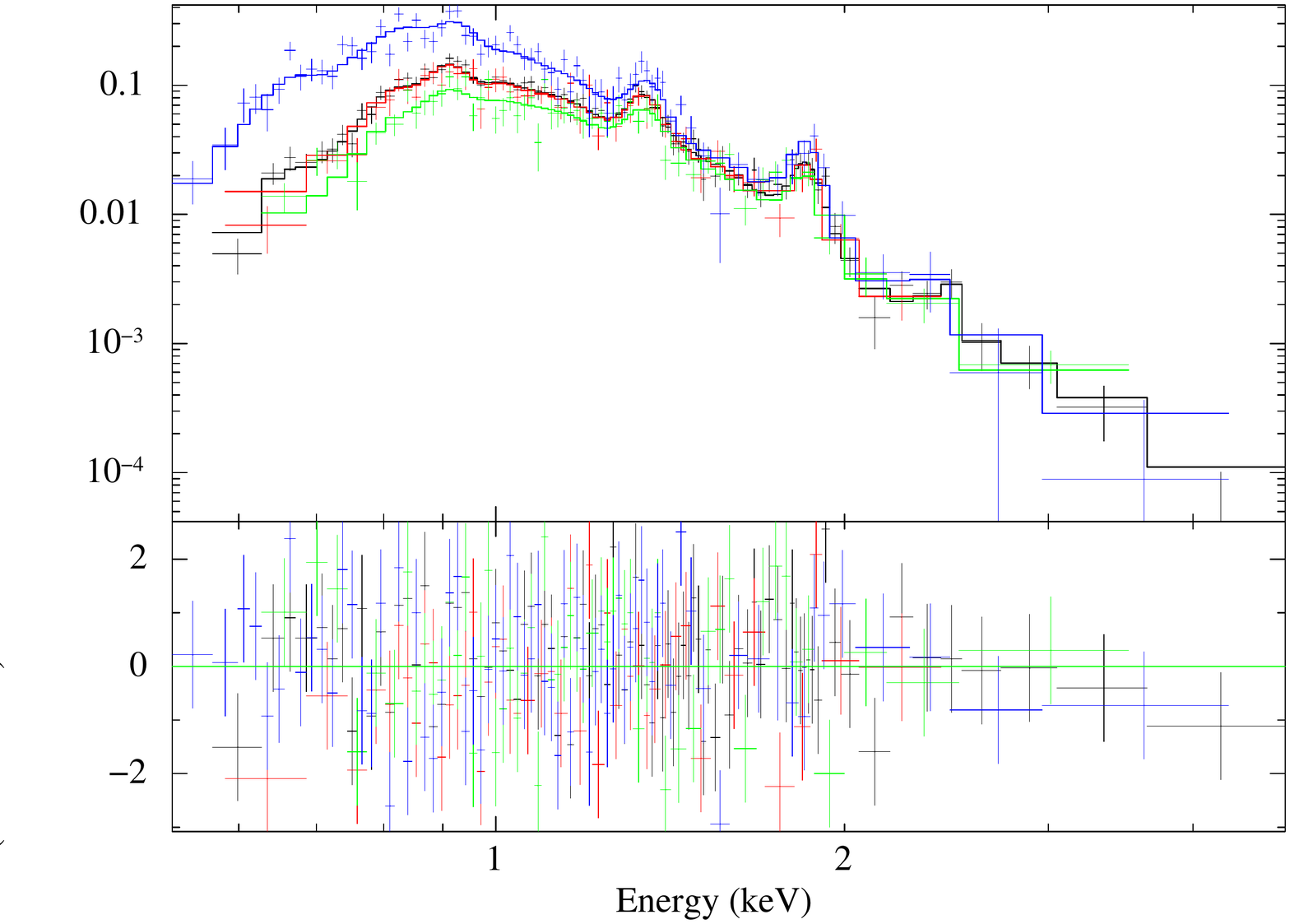}}
		\subfloat[(b) Region 7 (VAPEC)]{\includegraphics[angle=0,width=0.38\textwidth]{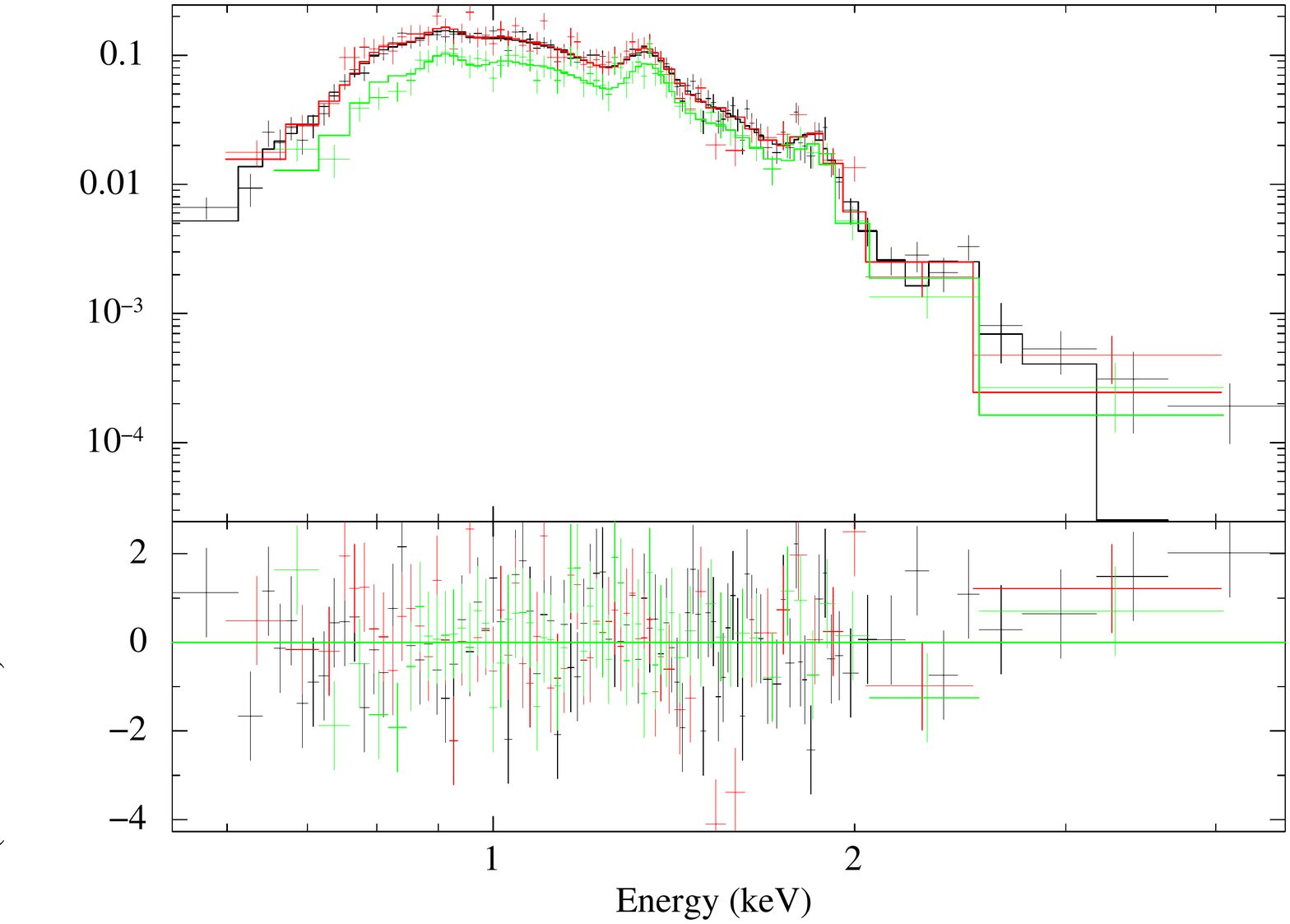}}
		\\
		\subfloat[(c) Region 8 (VPSHOCK)]{\includegraphics[angle=0,width=0.38\textwidth]{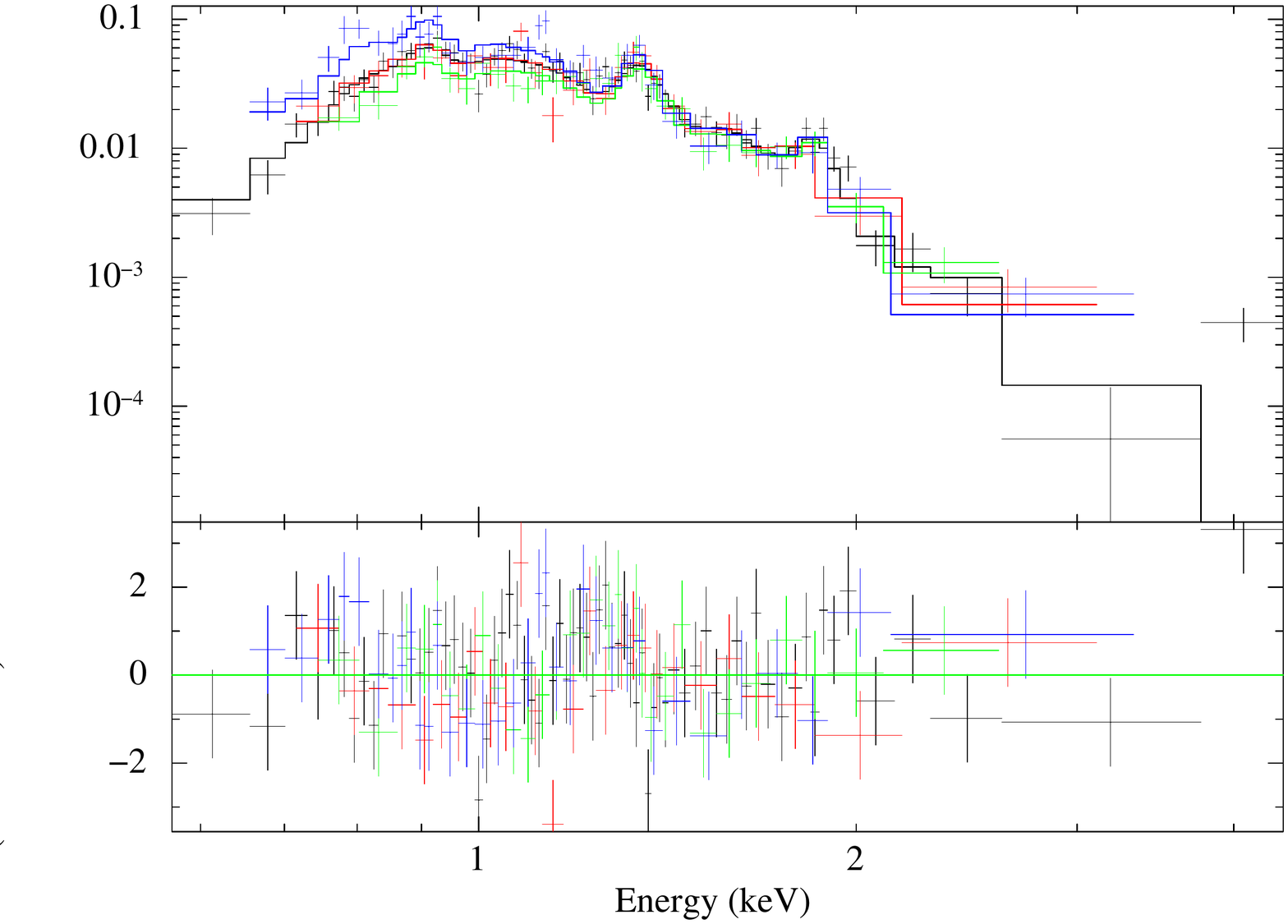}}
		\subfloat[(d) Region 18 (VPSHOCK)]{\includegraphics[angle=0,width=0.38\textwidth]{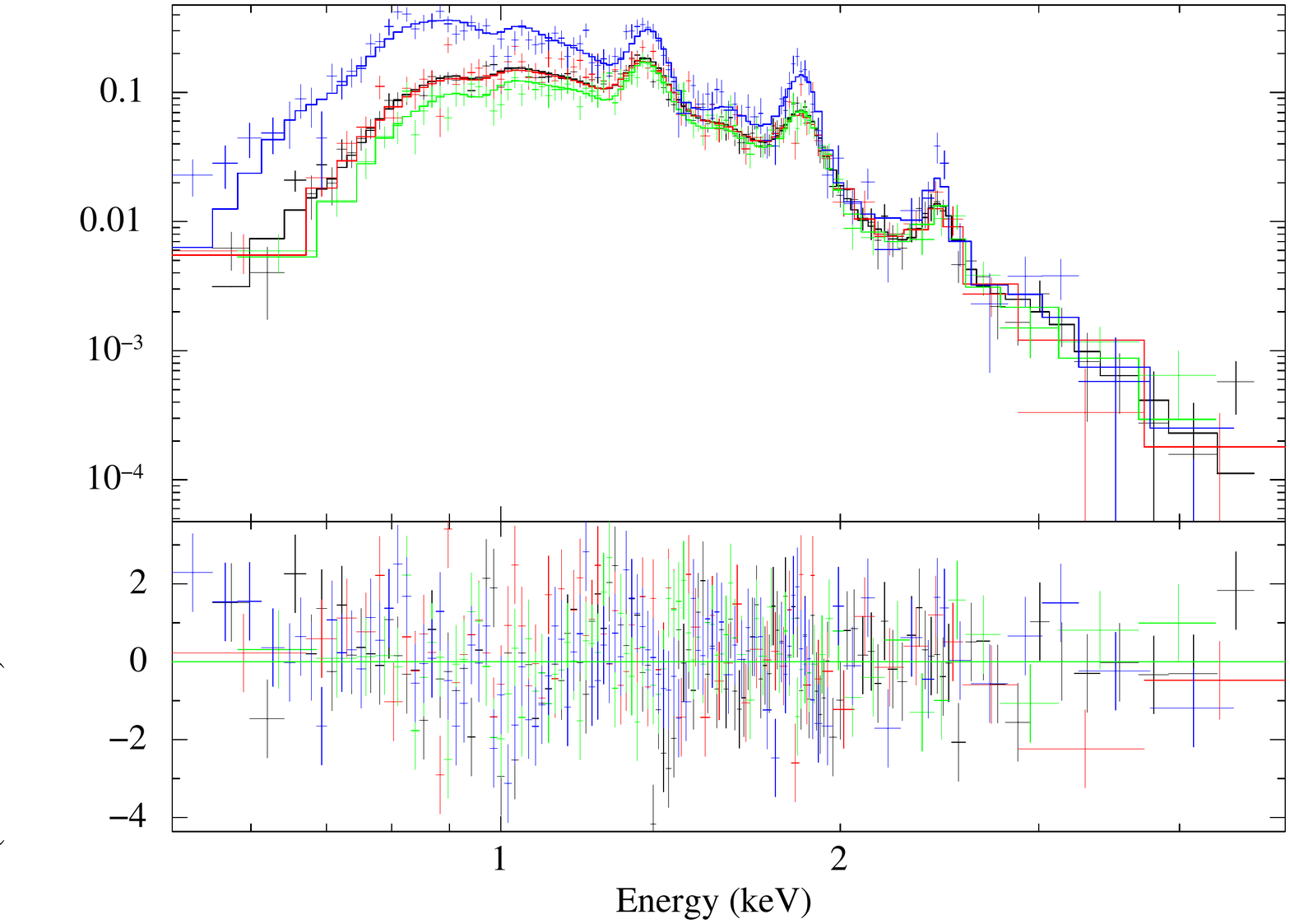}}
		\\
		\subfloat[(e) Region 33a (VPSHOCK+APEC; South Limb)]{\includegraphics[angle=0,width=0.38\textwidth]{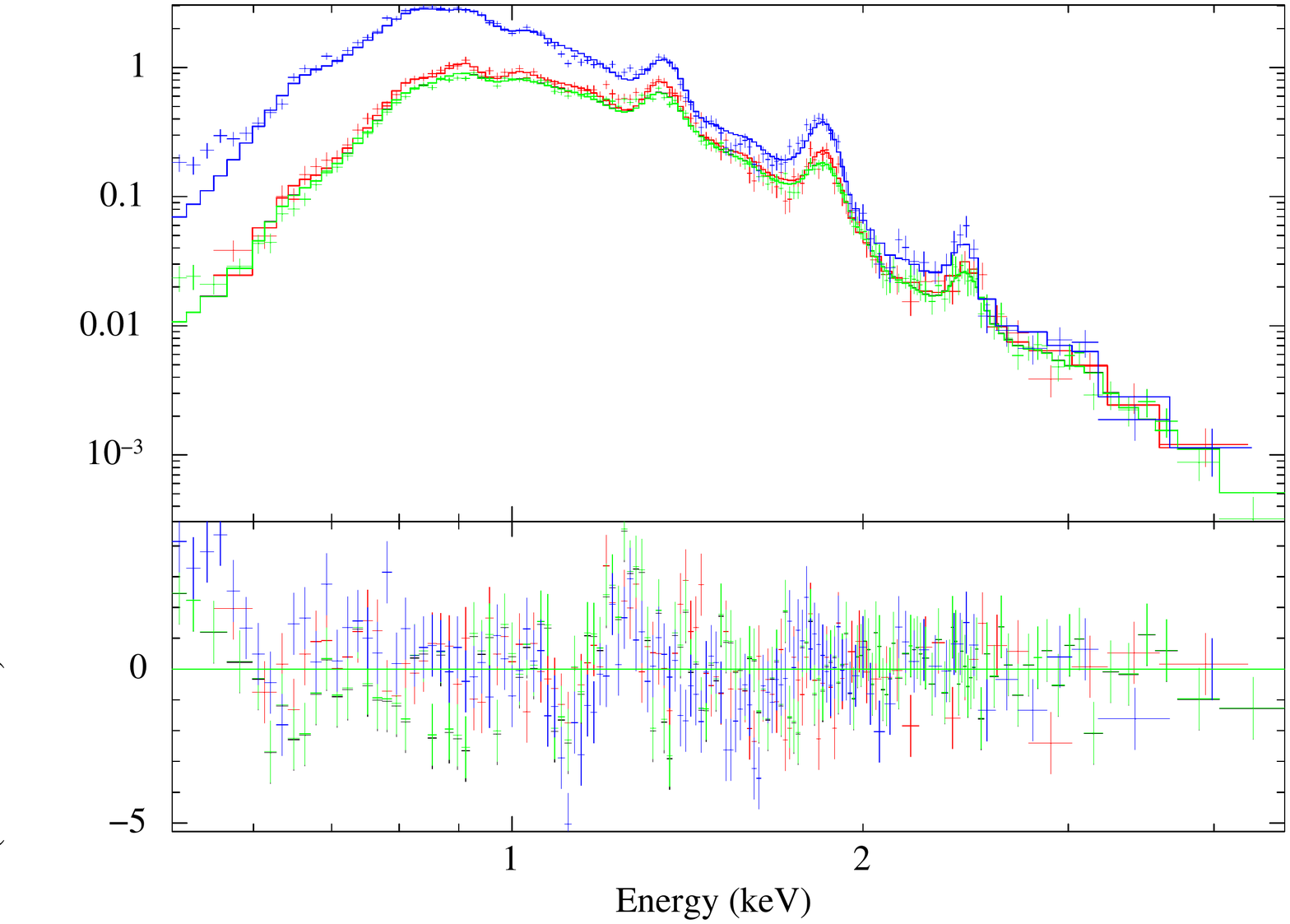}}
		\subfloat[(f) Region 21b (VPSHOCK+APEC; North Limb)]{\includegraphics[angle=0,width=0.38\textwidth]{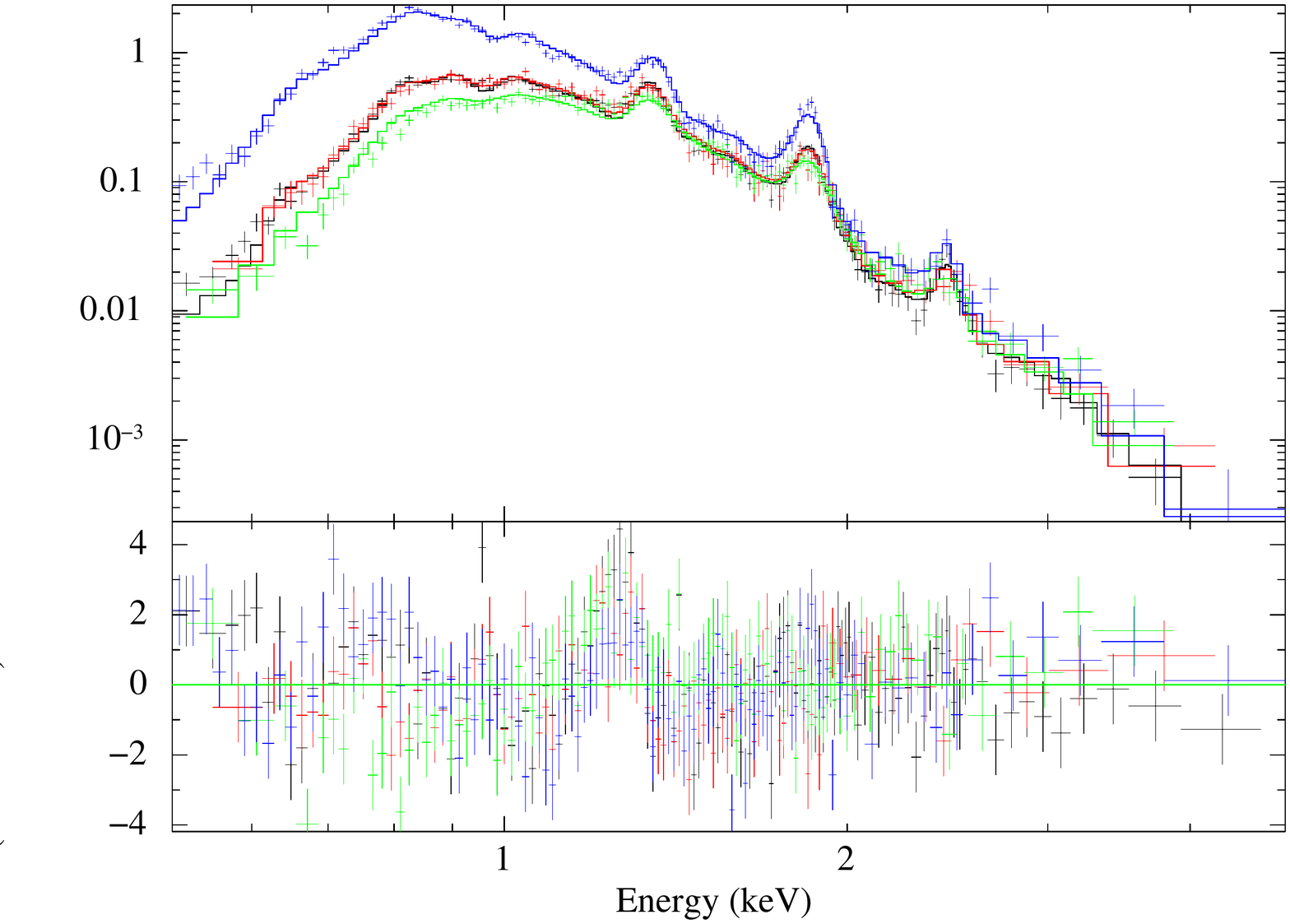}}

		\caption{Example spectra from Table~\ref{tbl:OneCompData} or \ref{tbl:RegionData} with energy ranges from 0.5--5.0~keV. Each graph has normalized counts vs energy (top) and residual plots (bottom). }
		\label{fig:Spectra}	
		\end{center}
	\end{figure*}

    Each region was first fit with a single VPSHOCK model, followed by a two-component VPSHOCK+VPSHOCK model (as required by the data), but found that the soft component required a CIE model. From there, a two-component VPSHOCK+APEC was fit, and the most statistically significant fit (based on the reduced chi-squared statistics, $\chi^2_\nu$), either the one or two-component models, were recorded into Tables~\ref{tbl:OneCompData} or \ref{tbl:RegionData}, respectively. Example spectra can be found in Fig.~\ref{fig:Spectra}. Uncertainties in the measurements were calculated using the `steppar' XSPEC command, which creates contour plots by stepping incrementally through a two-parameter space and performing fits. Each error listed is at the $2\sigma$.  
		
	\subsubsection{One-Component Models}
	
	The regions were first fit with a single component model VPSHOCK. If the one-component model fits reached the maximum ionization timescale, n$_\text{e}\text{t} = 5 \times 10^{13}$~cm$^{-3}$~s, it was fit with a single VAPEC model, which allows for individually varying abundances (unlike the APEC model) and corresponds to plasma in CIE. The fitting process begins with all abundance variables frozen at solar, and only allowing the constants, column density, temperature, normalization factor, and ionization timescale, if applicable, to vary. From there, the abundance variables were freed and fit one at a time from Mg, Si, S, and Fe tethered to Ni. See Table~\ref{tbl:OneCompData}. Generally regions that were smaller or had lower count rates were best fit by the single component models ($\chi^2_\nu < 2$), however, each region was also fit with a two-component model and the most statistically significant model was used. We also calculated the F-statistic between the one- and two-component fits to determine if the secondary component was statistically justified. The regions can be separated into hard and soft components that mimic the two-component models based on their temperature and whether they are in NEI or CIE (see Section 4.2). We classified the hard component regions by fitting with a VPSHOCK model, had temperatures greater than 0.38~keV, and were still in NEI. The soft component regions were fit with a VAPEC model, had temperatures below 0.38~keV and were in CIE, of which there was only three. However, the hard component of the two-component regions tended to be hotter, $\sim0.6$~keV. Abundance values seemed roughly solar or slightly subsolar for most regions except in some of the soft component models which had some supersolar values.
	
	\begin{table*} 
			
		\caption{Spectral properties of selected regions fitted with one-component models. Models were considered hard if the temperature was higher than 0.38~keV and with an ionization timescale in the range of $10^{11}$--$10^{12}$~cm$^{-3}$~s and soft if the temperatures were below 0.38~keV and in CIE. Regions were fit with the VPSHOCK model except for regions 7, 10, and 14 which were fit with the VAPEC model.}
			
		\label{tbl:OneCompData} 
		\begin{tabular}{ccccccccc}
		\hline
		
		Region & N$_\text{H}$ 						& kT 	& Mg & Si &	S &	Fe (Ni) & n$_\text{e}$t 					   & $\chi^2_\nu$ (DOF) \\   
     		   & $\times 10^{22}$~cm$^{-2}$	& keV &    &    &   &  	 	& $\times 10^{11}$~cm$^{-3}$~s & \\ 	
		\hline

		5 & $1.1^{+0.1}_{-0.1}$ & $0.42^{+0.05}_{-0.08}$ & $0.8^{+0.4}_{-0.3}$ & $0.7^{+0.3}_{-0.3}$ & $0.6^{+0.4}_{-0.3}$ & $0.8^{+0.3}_{-0.2}$ & $2.1^{+3.4}_{-1.1}$  & 0.99 (191) \\
        6 & $0.74^{+0.08}_{-0.07}$ & $0.69^{+0.09}_{-0.06}$ & $1.05^{+0.01}_{-0.47}$ & $0.80 (< 0.95)$ & $0.44 (< 0.97) $ & $1.0^{+0.3}_{-0.2}$ & $2.1^{+3.4}_{-0.9}$  & 1.12 (250) \\
        7 & $1.26^{+0.12}_{-0.15}$ & $0.24^{+0.04}_{-0.02}$ & $1.0^{+0.2}_{-0.2}$ & $2.0^{+1.4}_{-0.7}$ & $10.4^{+7.1}_{-5.0}$ & $1.1^{+0.2}_{-0.1}$ & ...  & 1.14 (209) \\
        8 & $1.3^{+0.2}_{-0.1}$ & $0.49^{+0.10}_{-0.09}$ & $0.7^{+0.4}_{-0.3}$ & $0.9^{+0.2}_{-0.2}$ & ... & $0.6^{+0.3}_{-0.2}$ & $2.5^{+6.0}_{-1.6}$  & 1.19 (155) \\
        10 & $1.1^{+0.1}_{-0.1}$ & $0.29^{+0.02}_{-0.03}$ & $0.8^{+0.2}_{-0.1}$ & $1.5^{+0.6}_{-0.7}$ & $4.2^{+6.3}_{-2.1}$ & $0.8^{+0.2}_{-0.2}$ & ... & 1.00 (214) \\
        12 & $1.2^{+0.1}_{-0.1}$ & $0.38^{+0.10}_{-0.08}$ & $0.8^{+0.1}_{-0.1}$ & $1.1^{+0.5}_{-0.3}$ & $1.2^{+2.0}_{-1.0}$ & $0.6^{+0.2}_{-0.1}$ & $4.9^{+10.6}_{-2.4}$ & 1.12 (289) \\
        14 & $1.30^{+0.15}_{-0.07}$ & $0.24^{+0.03}_{-0.07}$ & $1.2^{+2.4}_{-0.1}$ & $3.2^{+17.1}_{-0.4}$ & $16.0 (> 11.4)$ & $1.4^{+3.4}_{-0.2}$ & ...  & 1.59 (389) \\
        18 & $1.25^{+0.1}_{-0.09}$ & $0.60^{+0.08}_{-0.06}$ & $1.3^{+0.3}_{-0.2}$ & $1.6^{+0.3}_{-0.3}$ & $1.7^{+0.5}_{-0.5}$ & $0.9^{+0.3}_{-0.3}$ & $2.2^{+1.5}_{-0.8}$ & 1.36 (374) \\
        44 & $1.22^{+0.02}_{-0.13}$ & $0.70^{+0.04}_{-0.12}$ & $2.3^{+0.1}_{-0.8}$ & $2.7^{+0.1}_{-1.0}$ & $1.5^{+0.7}_{-0.4}$ & $1.8^{+0.4}_{-0.7}$ & $2.0^{+1.8}_{-0.5}$ & 1.20 (251) \\
        45 & $1.40^{+0.06}_{-0.07}$ & $0.60^{+0.08}_{-0.03}$ & $1.5^{+0.5}_{-0.3}$ & $1.7^{+0.4}_{-0.2}$ & $1.0^{+0.4}_{-0.3}$ & $1.2^{+0.7}_{-0.3}$ & $2.7^{+1.2}_{-0.8}$ &  1.28 (375) \\
        46 & $1.14^{+0.07}_{-0.04}$ & $0.55^{+0.05}_{-0.03}$ & $1.4^{+0.8}_{-0.3}$ & $1.5^{+1.0}_{-0.3}$ & $1.0^{+0.2}_{-0.3}$ & $1.0^{+0.2}_{-0.2}$ & $3.2^{+2.4}_{-1.8}$ & 1.36 (364) \\
        Bullet 1 & $0.9^{+0.2}_{-0.2}$ & $0.4^{+0.2}_{-0.2}$ & $0.6^{+0.5}_{-0.3}$ & $1.9^{+1.1}_{-3.2}$ & ... & $0.3^{+1.2}_{-0.1}$ & $7.7$ ($>1.3$) & 0.91 (105) \\
        Bullet 2 & $0.13^{+0.10}_{-0.08}$ & $0.81^{+0.05}_{-0.05}$ & $1.0^{+0.8}_{-0.5}$ & $0.7^{+0.7}_{-0.6}$ & ... & $0.5^{+0.2}_{-0.1}$ & ... & 1.37 (105) \\
        Bullet 3 & $0.5^{+0.2}_{-0.1}$ & $0.9^{+0.8}_{-0.2}$ & $1.7^{+1.0}_{-0.7}$ & $1.1^{+0.2}_{-0.6}$ & ... & $1.0^{+1.2}_{-0.5}$ & $1.6^{+2.7}_{-0.6}$ & 1.07 (139) \\
        Bullet 4 & $1.0^{+0.2}_{-0.2}$ & $0.7^{+0.4}_{-0.2}$ & $1.5$ ($>0.4$) & $1.4$ ($>0.8$) & $0.9^{+1.6}_{-0.9}$ & $1.4^{+2.5}_{-0.9}$ & $2.5^{+4.9}_{-1.6}$ & 1.25 (103) \\
        Bullet 5 & $1.0^{+0.2}_{-0.4}$ & $0.4^{+0.4}_{-0.1}$ & $0.8^{+0.9}_{-0.2}$ & $1.4^{+2.3}_{-0.7}$ & ... & $0.5^{+0.3}_{-0.2}$ & $5.4$ ($>3.6$) & 1.35 (98) \\
		
		\hline 	
		\end{tabular}
			
	\end{table*}	

	\subsubsection{Two-Component Models}
	
	For a majority of the regions, a single component model was not statistically successful ($\chi^2_\nu > 2$) and so a secondary component was required to account for any mixing of shocked ejecta and circumstellar material. This was motivated by the failure of the one-component models and with the expectation of a high- and low-temperature plasma associated with the supernova blast wave and reverse-shocked ejecta as seen in many SNRs (e.g. \cite{Samar2005, 2014ApJ...781...41K}). The two-component regions were fit with a VPSHOCK+APEC model and can be found in Table~\ref{tbl:RegionData}.	All small-scale regions	were well fit with $\chi^2_\nu<2$, where imperfect fits are likely due to non-uniform spectral properties within a single region requiring additional components. However, given the data quality and the large number of parameters to fit, we opted for a two-component fit which would account for the shock-heated ISM/CSM and ejecta components. We note however that our fits are adequate with $\chi^2_\nu<2$,and such quality fits have been adopted for several \textit{Chandra}-studied ejecta-dominated SNRs (e.g. \cite{HwangPuppis, ParkG299, Samar2005}).

    The fitting process involved freezing the abundance variables at solar, and allowing the constants, temperatures, normalization factor, and ionization timescale to vary. From there, the abundances were freed and fit one at a time from Mg, Si, S, to Fe tethered to Ni. Originally the regions were fit with a VPSHOCK+PSHOCK model, but the secondary component's ionization timescale always reached the maximum allowed value and implied that this component was in CIE, and hence fit with the APEC model. It should be noted, that both the hard and soft components abundances were varied while the other component was held at solar, but the most statistically significant results allowed the hard component abundances to vary and the soft component abundances frozen at solar. The hard component associated with the VPSHOCK model had higher temperatures, $\sim0.60$~keV, ionization timescales around $10^{11}$--$10^{12}$~cm$^{-3}$~s, and variable abundances that range from solar to slightly supersolar. The soft component associated with the APEC model had temperatures $\sim0.2$~keV and abundances frozen at solar. The hydrogen column density ranges from $(0.86$--$1.46) \times 10^{22}$~cm$^{-2}$. To see a discussion on the global trends see the next section.
	
	\begin{table*}
	    \caption{Spectral properties of selected regions fitted with the two-component VPSHOCK+APEC model. The VPSHOCK component has varied Mg, Si, S, and Fe (= Ni) abundances and the APEC model component has abundances frozen at solar values. If an abundance value is missing, the value was frozen at solar. For Reg29a, we freed and fit the Neon abundance to get an acceptable fit  (with $\chi^2_\nu<2$), where $Ne=2.2^{+0.6}_{-0.3}$. Errors are $2\sigma$ confidence level,
	    except for the full SNR which has no errors reported. For the latter global fit, we only show representative spectral values.}
			
		\label{tbl:RegionData} 
		\begin{tabular}{cccccccccc}
		\hline
		Region & N$_\text{H}$ 						& kT$_h$ 	& Mg & Si &	S &	Fe (Ni) & n$_\text{e}$t 					   & kT$_s$    & $\chi^2_\nu$ (DOF) \\  		  
		       & $\times 10^{22}$~cm$^{-2}$	& keV &    &    &   &  	 	    & $\times 10^{11}$~cm$^{-3}$~s & keV & 			        \\ 
		\hline
		1 & $1.02^{+0.09}_{-0.2}$ & $0.59^{+0.3}_{-0.08}$ & $2.0^{+0.8}_{-0.6}$ & $2.0^{+0.8}_{-0.5}$ & $1.6^{+1.4}_{-1.0}$ & $1.9^{+1.2}_{-0.7}$ & $2.6^{+3.6}_{-1.4}$ & $0.22^{+0.09}_{-0.18}$ & 1.47 (175) \\
        2 & $1.18^{+.09}_{-0.12}$ & $0.59^{+0.14}_{-0.09}$ & $1.1^{+0.6}_{-0.4}$ & $1.4^{+0.7}_{-0.4}$ & $1.0^{+1.2}_{-0.7}$ & $1.4^{+0.8}_{-0.5}$ & $3.0^{+2.2}_{-2.1}$ & $0.19^{+0.05}_{-0.04}$ & 1.06 (278) \\
        3 & $1.25\pm{0.05}$ & $0.55\pm{0.04}$ & $1.4^{+4.5}_{-0.4}$ & $1.8^{+9.4}_{-0.3}$ & $1.4^{+1.2}_{-0.6}$ & $1.7^{+0.8}_{-0.2}$ & $4.2^{+7.7}_{-2.8}$ &  $0.205^{+0.009}_{-0.007}$ & 1.38 (295) \\
        4 & $1.3^{+0.1}_{-0.2}$ & $0.6\pm{0.2}$ & $1.9^{+1.6}_{-0.7}$ & $1.8^{+2.0}_{-0.6}$ & $1.5^{+2.8}_{-1.1}$ & $1.7^{+1.7}_{-0.8}$ & $5.0 (> 3.4)$ & $0.21^{+0.05}_{-0.06}$ & 1.12 (222) \\
        9 & $1.34^{+0.09}_{-0.08}$ & $0.54^{+0.07}_{-0.08}$ & $1.8^{+0.7}_{-0.4}$ & $2.1^{+0.9}_{-0.5}$ & $1.2^{+1.0}_{-0.5}$ & $1.8^{+1.0}_{-0.5}$ & $4.9^{+8.8}_{-2.2}$ & $0.22^{+0.03}_{-0.05}$ & 1.12 (247) \\
        11 & $1.2\pm{0.1}$ & $0.68^{+0.04}_{-0.03}$ & ... & $2.0^{+2.2}_{-0.5}$ & $1.6^{+1.2}_{-0.3}$ & $1.7^{+0.6}_{-1.0}$ & 3.7~$(>1.0)$ & $0.24^{+0.04}_{-0.01}$ & 1.10 (284) \\
        13 & 1.46$\pm{0.08}$ & $0.5^{+0.2}_{-0.1}$ & $1.3^{+0.7}_{-0.4}$ & $1.8^{+1.2}_{-0.5}$ &  $1.4^{+1.5}_{-0.7}$ & $1.6^{+1.3}_{-0.5}$ & $4.8^{+31.0}_{-2.9}$ & $0.19^{+0.05}_{-0.03}$ & 1.15 (322) \\
        15 & $1.09\pm{0.04}$ & $0.60^{+0.07}_{-0.06}$ & $1.6^{+0.4}_{-0.3}$ & $2.0^{+0.5}_{-0.4}$ & $1.2\pm{0.5}$ & $1.7^{+0.5}_{-0.3}$ & $4.8^{+3.5}_{-1.8}$ & $0.23\pm{0.02}$ & 1.55 (416) \\
        16 & $1.1\pm{0.1}$ & $0.9^{+0.4}_{-0.2}$ & $1.2^{+0.5}_{-1.2}$ & $2.0^{+2.1}_{-0.6}$ & $1.1^{+2.5}_{-0.9}$ & $2.0^{+2.8}_{-0.8}$ & $2.1^{+3.1}_{-1.3}$ & $0.22^{+0.02}_{-0.03}$ & 1.13 (257) \\
        17 & $1.44^{+0.07}_{-0.08}$ & $0.52^{+0.05}_{-0.07}$ & $1.6^{+0.4}_{-0.3}$ & $1.9^{+0.5}_{-0.3}$ & $1.6^{+0.7}_{-0.5}$ & $1.7^{+0.5}_{-0.4}$ & $4.5^{+2.8}_{-1.5}$ & $0.18^{+0.03}_{-0.04}$ & 1.42 (400) \\
        19 & $1.42\pm{0.07}$ & $0.60^{+0.1}_{-0.6}$ & $1.4\pm{0.2}$ & $2.0^{+0.4}_{-0.2}$ & $2.1^{+0.9}_{-0.6}$ & $1.1^{+0.5}_{-0.4}$ & $3.6^{+4.3}_{-2.1}$ & $0.21\pm{0.03}$ & 1.31 (372) \\
        20 & $1.1\pm{0.1}$ & $0.7^{+0.5}_{-0.2}$ & $1.7^{+1.0}_{-0.4}$ & $1.7^{+1.9}_{-0.4}$ & $0.9^{+0.8}_{-0.5}$ & $2.1^{+0.8}_{-0.4}$ & $2.2^{+1.5}_{-1.6}$ & $0.19^{+0.06}_{-0.05}$ & 1.14 (289) \\
        21a & $1.09\pm{0.04}$ & $0.55^{+0.04}_{-0.03}$ & $1.7\pm{0.3}$ & $1.8^{+0.4}_{-0.2}$ & $1.2^{+0.5}_{-0.4}$ & $1.8^{+0.4}_{-0.3}$ & $7.1^{+3.7}_{-2.2}$ & $0.20^{+0.3}_{-0.2}$ & 1.48 (418) \\
        21b & 1.13$^{+0.04}_{-0.08}$ & $0.51^{+0.04}_{-0.03}$ & $1.3^{+0.2}_{-0.1}$ & $1.7\pm{0.2}$ & $1.1\pm{0.3}$ & $1.3^{+0.2}_{-0.1}$ & $14^{+24}_{-5}$ & $0.21^{+0.01}_{-0.02}$ & 1.71 (453) \\
        22 & $1.2\pm{0.1}$ & $0.54^{+0.09}_{-0.07}$ & $1.3^{+0.6}_{-0.3}$ & $1.5^{+0.7}_{-0.4}$ & $1.2^{+1.0}_{-0.7}$ & $1.5^{+0.7}_{-0.4}$ & $7.1^{+23}_{-3.7}$ & $0.22^{+0.06}_{-0.05}$ & 1.16 (215) \\
        23 & $0.96^{+0.30}_{-0.07}$ & $0.54^{+0.15}_{-0.06}$ & $1.4^{+0.5}_{-0.3}$ & $1.8^{+0.9}_{-0.4}$ & $1.3^{+1.0}_{-0.7}$ & $1.5^{+0.7}_{-0.4}$ & $4.2^{+5.4}_{-2.8}$ & $0.22^{+0.03}_{-0.03}$ & 1.26 (258) \\
        24 & $0.97^{+0.05}_{-0.06}$ & $0.55^{+0.04}_{-0.06}$ & $1.3^{+0.6}_{-0.3}$ & $1.4^{+0.5}_{-0.4}$ & $1.1\pm{0.3}$ & $1.4^{+0.5}_{-0.4}$ & $4.5^{+3.8}_{-1.4}$ & $0.18^{+0.05}_{-0.03}$ & 1.99 (489) \\
        25a & $1.1^{+0.1}_{-0.2}$ & $0.6\pm{0.2}$ & $1.5^{+0.9}_{-0.5}$ & $1.0^{+0.7}_{-0.4}$ & $1.0^{+2.0}_{-0.8}$ & $1.6^{+0.9}_{-0.7}$ & $5.9 (> 1.9)$ & $0.18^{+0.06}_{-0.04}$ & 1.07 (242) \\
        25b & $1.12^{+0.08}_{-0.07}$ & $0.54^{+0.04}_{-0.05}$ &  $1.1^{+0.4}_{-0.3}$ & $1.3^{+1.3}_{-0.3}$ & $1.8^{+6.2}_{-0.8}$ & $0.9^{+3.4}_{-0.4}$ & $4.2 (> 1.4)$ & $0.22\pm{0.03}$ & 1.15 (290) \\
        26 & $1.1\pm{0.1}$ & $0.62^{+0.06}_{-0.05}$ & $1.3\pm{0.2}$ & $1.2\pm{0.2}$ & $1.0^{+0.4}_{-0.3}$ & $1.0^{+0.3}_{-0.2}$ & $5.2^{+2.8}_{-1.6}$ & $0.18^{+0.04}_{-0.05}$ & 1.28 (391) \\
        27 & $0.87\pm{0.07}$ & $0.54^{+0.04}_{-0.05}$ & $1.6\pm{0.2}$ & $1.6^{+0.4}_{-0.3}$ & $0.9^{+0.5}_{-0.4}$ & $1.6^{+0.4}_{-0.3}$ & $4.4^{+2.3}_{-1.2}$ & $0.19\pm{0.01}$ &  1.35 (351) \\
        28a & $0.95^{+0.08}_{-0.04}$ & $0.55^{+0.04}_{-0.02}$ & $1.3\pm{0.1}$ & $1.5\pm{0.1}$ & $1.1^{+1.4}_{-0.6}$ & $1.3^{+1.2}_{-0.7}$ & $6.0^{+0.4}_{-1.3}$ & $0.187^{+0.009}_{-0.006}$ &  1.86 (498) \\
        28b & $0.87^{+0.05}_{-0.04}$ & $0.53^{+0.03}_{-0.05}$ & $1.5\pm{0.2}$ & $1.5^{+0.3}_{-0.2}$ & $0.9^{+0.5}_{-0.3}$ & $1.6\pm{0.3}$ & $6.3^{+5.4}_{-1.6}$ & $0.185^{+0.009}_{-0.006}$ &  1.45 (403) \\
        29a & $0.97^{+0.07}_{-0.06}$ & $0.68^{+0.06}_{-0.04}$ & $2.8^{+0.7}_{-0.3}$ & $2.0^{+0.5}_{-0.3}$ & $1.1\pm{0.4}$ & $1.7\pm{0.3}$ & $4.0^{+1.4}_{-1.2}$ & $0.20^{+0.03}_{-0.02}$ & 1.90 (420) \\
        29b & $1.06^{+0.5}_{-0.04}$ & $0.54^{+0.06}_{-0.07}$ & $1.3^{+0.4}_{-0.3}$ & $1.5\pm{0.4}$ & $1.0^{+0.3}_{-0.4}$ & $1.2\pm{0.2}$ & $6.1^{+4.2}_{-2.3}$ & $0.18^{+0.07}_{-0.05}$ & 1.31 (348) \\
        30a & $0.99\pm{0.06}$ & $0.55^{+0.02}_{-0.03}$ & $1.1^{+0.2}_{-0.1}$ & $1.1^{+0.2}_{-0.5}$ & $0.8^{+0.4}_{-0.2}$ & $1.1^{+0.3}_{-0.1}$ & $8.1^{+0.4}_{-2.0}$ & $0.18^{+0.01}_{-0.02}$ & 1.52 (432) \\
        30b & $0.99\pm{0.06}$ & $0.55\pm{0.02}$ & $1.1^{+0.2}_{-0.1}$ & $0.8\pm{0.2}$ & $0.8\pm{0.2}$ & $1.1^{+0.3}_{-0.1}$ & $8.1^{+0.4}_{-2.0}$ & $0.18^{+0.01}_{-0.03}$ &  1.52 (420) \\
        33a & $1.10^{+0.03}_{-0.04}$ & $0.71^{+0.05}_{-0.03}$ & $1.3\pm{0.2}$ & $1.5^{+0.4}_{-0.2}$ & $1.0^{+0.3}_{-0.2}$ & $1.6^{+0.4}_{-0.2}$ & $3.6^{+0.7}_{-0.8}$ & $0.25\pm{0.01}$ &  1.80 (516) \\
        33b & $0.93\pm{0.05}$ & $0.57^{+0.03}_{-0.02}$ & $0.95\pm{0.09}$ & $0.85^{+0.09}_{-0.07}$ & $0.6^{+0.2}_{-0.1}$ & $0.9\pm{0.1}$ & $3.6^{+1.3}_{-0.8}$ & $0.18^{+0.01}_{-0.05}$ &  1.95 (513) \\
        34 & $1.33\pm{0.05}$ & $0.58^{+0.12}_{-0.04}$ & $1.7\pm{0.3}$ & $1.5\pm{0.2}$ & $0.9\pm{0.3}$ & $1.4^{+0.4}_{-0.3}$ & $4\pm{2}$ & $0.19^{+0.04}_{-0.01}$ &  1.46 (441) \\
        35 & $0.94\pm{0.04}$ & $0.66^{+0.05}_{-0.06}$ & $1.6^{+0.7}_{-0.6}$ & $1.6^{+0.4}_{-0.5}$ & $1.1\pm{0.2}$ & $1.7^{+0.5}_{-0.4}$ & $3.5^{+2.2}_{-1.8}$ & $0.24\pm{0.02}$ &  1.92 (480) \\
        36a & $1.03\pm{0.03}$ & $0.78^{+0.08}_{-0.05}$ & $2.5^{+0.7}_{-0.6}$ & $3.0^{+1.0}_{-0.7}$ & $1.7^{+1.1}_{-0.3}$ & $2.8^{+0.9}_{-0.8}$ & $2.4^{+0.9}_{-0.8}$ & $0.28\pm{0.01}$ &  1.89 (438) \\
        36b & $1.20\pm{0.07}$ & $0.46^{+0.06}_{-0.04}$ & $1.5\pm{0.2}$ & $1.7^{+0.4}_{-0.2}$ & $1.5^{+0.7}_{-0.4}$ & $1.2^{+0.4}_{-0.2}$ & $0.18^{+0.02}_{-0.03}$ & $10.4^{+2.7}_{-6.9}$ & 1.51 (421) \\
        37a & $1.11^{+0.03}_{-0.05}$ & $0.56^{+0.03}_{-0.04}$ & $1.2\pm{0.2}$ & $1.5\pm{0.2}$ & $0.8^{+0.2}_{-0.3}$ & $1.5^{+0.2}_{-0.3}$ & $6.4^{+4.0}_{-1.8}$ & $0.18\pm{0.01}$ &  1.80 (443) \\
        37b & $1.12^{+0.08}_{-0.03}$ & $0.54^{+0.06}_{-0.09}$ & $1.3^{+0.4}_{-0.3}$ & $1.5^{+0.4}_{-0.3}$ & $1.3^{+0.7}_{-0.5}$ & $1.4^{+0.6}_{-0.3}$ & $4.1^{+9}_{-1.0}$ & $0.18^{+0.01}_{-0.03}$ &  1.35 (334) \\
        38a & $1.07^{+0.05}_{-0.08}$ & $0.55^{+0.04}_{-0.03}$ & $0.9^{+0.2}_{-0.1}$ & $1.1^{+0.2}_{-0.1}$ & $1.0\pm{0.2}$ & $1.0\pm{0.2}$ & $8\pm{2}$ & $0.19^{+0.01}_{-0.04}$ &  1.71 (424) \\
        38b & $1.04^{+0.06}_{-0.08}$ & $0.60^{+0.10}_{-0.04}$ & $1.2\pm{0.2}$ & $1.3^{+0.3}_{-0.2}$ & $1.0^{+0.3}_{-0.2}$ & $1.00^{+0.04}_{-0.19}$ & $0.24^{+0.06}_{-0.04}$ & $7.2^{+3.4}_{-2.1}$ & 1.78 (417) \\
        38c & $1.08\pm{0.08}$ & $0.54\pm{0.05}$ & $1.3\pm{0.2}$ & $1.4^{+0.4}_{-0.2}$ & $1.4^{+0.6}_{-0.5}$ & $1.1^{+0.3}_{-0.2}$ & $9^{+15}_{-1}$ & $0.21\pm{0.04}$ &  1.41 (343) \\
        39 & $1.2\pm{0.1}$ & $0.7^{+0.2}_{-0.1}$ & $1.5\pm{0.4}$ & $1.7^{+0.6}_{-0.4}$ & $1.2^{+0.7}_{-0.4}$ & $1.2^{+0.5}_{-0.3}$ & $3.4^{+2.3}_{-1.4}$ & $0.22^{+0.03}_{-0.04}$ &  1.43 (352) \\
        40 & $1.06^{+0.04}_{-0.05}$ & $0.54^{+0.02}_{-0.04}$ & $1.2^{+0.2}_{-0.1}$ & $1.5^{+0.2}_{-0.3}$ & $1.0^{+0.3}_{-0.2}$ & $1.2^{+0.1}_{-0.2}$ & $7.8^{+2.4}_{-1.8}$ & $0.21^{+0.03}_{-0.02}$ &  1.92 (478) \\
        41 & $1.01\pm{0.06}$ & $0.55\pm{0.04}$ & $1.3\pm{0.2}$ & $1.6^{+0.3}_{-0.2}$ & $0.9^{+0.4}_{-0.3}$ & $1.2^{+0.3}_{-0.2}$ & $4.6^{+2.3}_{-1.2}$ & $0.19^{+0.03}_{-0.02}$ &  1.61 (407) \\
        42 & $1.35^{+0.09}_{-0.12}$ & $0.9^{+0.4}_{-0.2}$ & $3.2^{+2.8}_{-0.9}$ & $2.5^{+2.6}_{-0.8}$ & $1.2^{+1.9}_{-0.8}$ & $3.4^{+4.2}_{-1.2}$ & $1.41^{+1.1}_{-0.8}$ & $0.25^{+0.05}_{-0.03}$ &  1.16 (305) \\
        43 & $1.3\pm{0.1}$ & $1.1^{+1.4}_{-0.4}$ & $3.1^{+1.7}_{-1.0}$ & $6.2^{+3.9}_{-1.8}$ & $6.4^{+14.1}_{-6.0}$ & $5.2^{+31}_{-2.3}$ & $0.14^{+0.32}_{-0.04}$ & $0.24^{+0.08}_{-0.03}$ & 1.20 (240) \\
        Full SNR & 1.05 & 0.56 & 1.3 & 1.4 & 1.0 & 1.2 & 6.1 & 0.19 & 12.2 (1154) \\
		\hline
		\end{tabular}
	\end{table*}

	\subsubsection{Global SNR Model and Global Trends}
	
	The full SNR fit was attempted  using several models including VPSHOCK, APEC, and VSEDOV. The VSEDOV model is based on the Sedov-Taylor model, another non-equilibrium ionization model, based on the Sedov-Taylor dynamics \citep{Sedov}. This model is characterized by the ionization timescale and the mean and electron temperatures immediately behind the shock. Attempts were made to obtain an adequate fit by varying the two temperatures separately or tethered together.	While neither yielded a statistically acceptable fit ($\chi^2_\nu > > 2$) we report the following spectral parameters which will be discussed later (see \S5.3): N$_\text{H}=1.1\times 10^{22}$~cm$^{-2}$, $\text{kT}=0.36$~keV, n$_\text{e}$t$=2.8\times10^{11}$~cm$^{-3}$~s, and abundances frozen at their solar values.	This model should account for the bulk of the emission from the blast wave component under the assumption of a Sedov evolutionary stage, and is not meant to account for the ejecta component. A two-component VPSHOCK+APEC model, adopted for our spatially resolved spectroscopy study, likewise does not provide an adequate fit to the global SNR fit, however it provides an average, representative result for the entire remnant that we also use for comparison to our spatially resolved spectroscopic study and for our discussion in $\S$5.3. The VPSHOCK+APEC global fit parameters are summarized in Table~\ref{tbl:RegionData}. 
	
	\begin{figure*}
		\begin{center}
		\subfloat[(a) N$_\text{H}$ ($\times 10^{22}$~cm$^{-2}$)]{\includegraphics[width=0.32\textwidth]{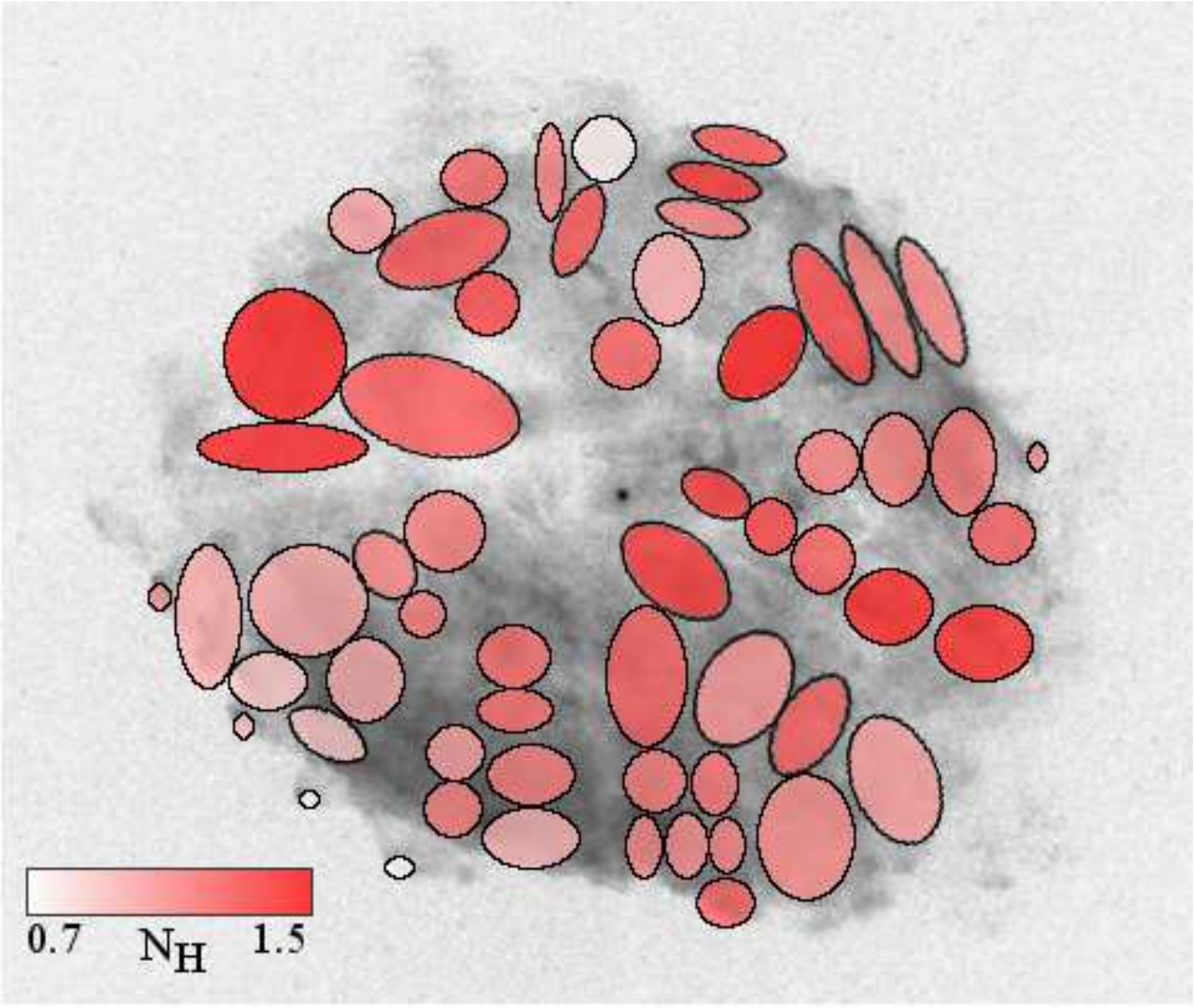}}
		\subfloat[(b) n$_\text{e}$t ($\times 10^{11}$~cm$^{-3}$~s)]{\includegraphics[width=0.32\textwidth]{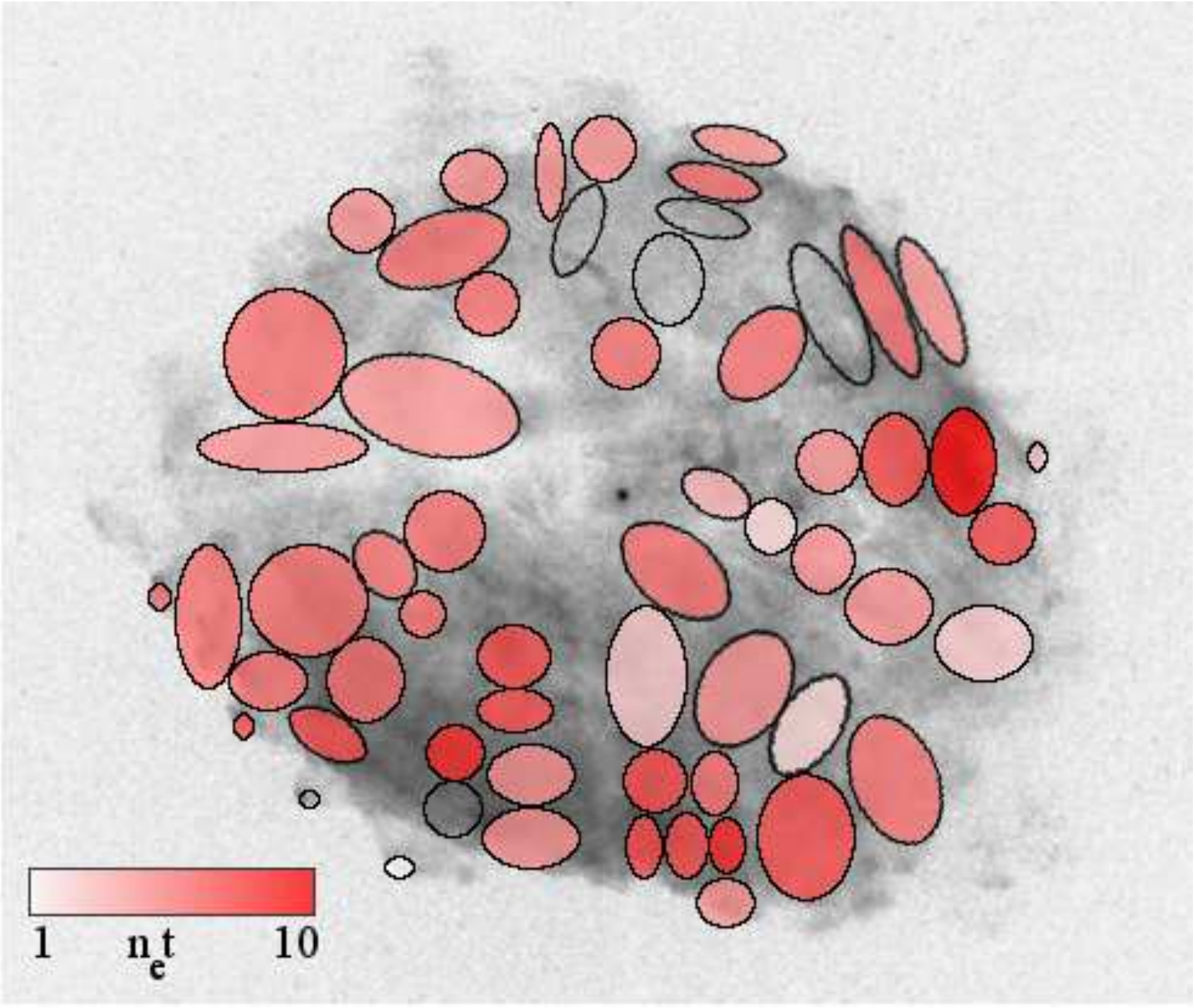}}
		\subfloat[(c) Temperature (keV; hard)]{\includegraphics[width=0.32\textwidth]{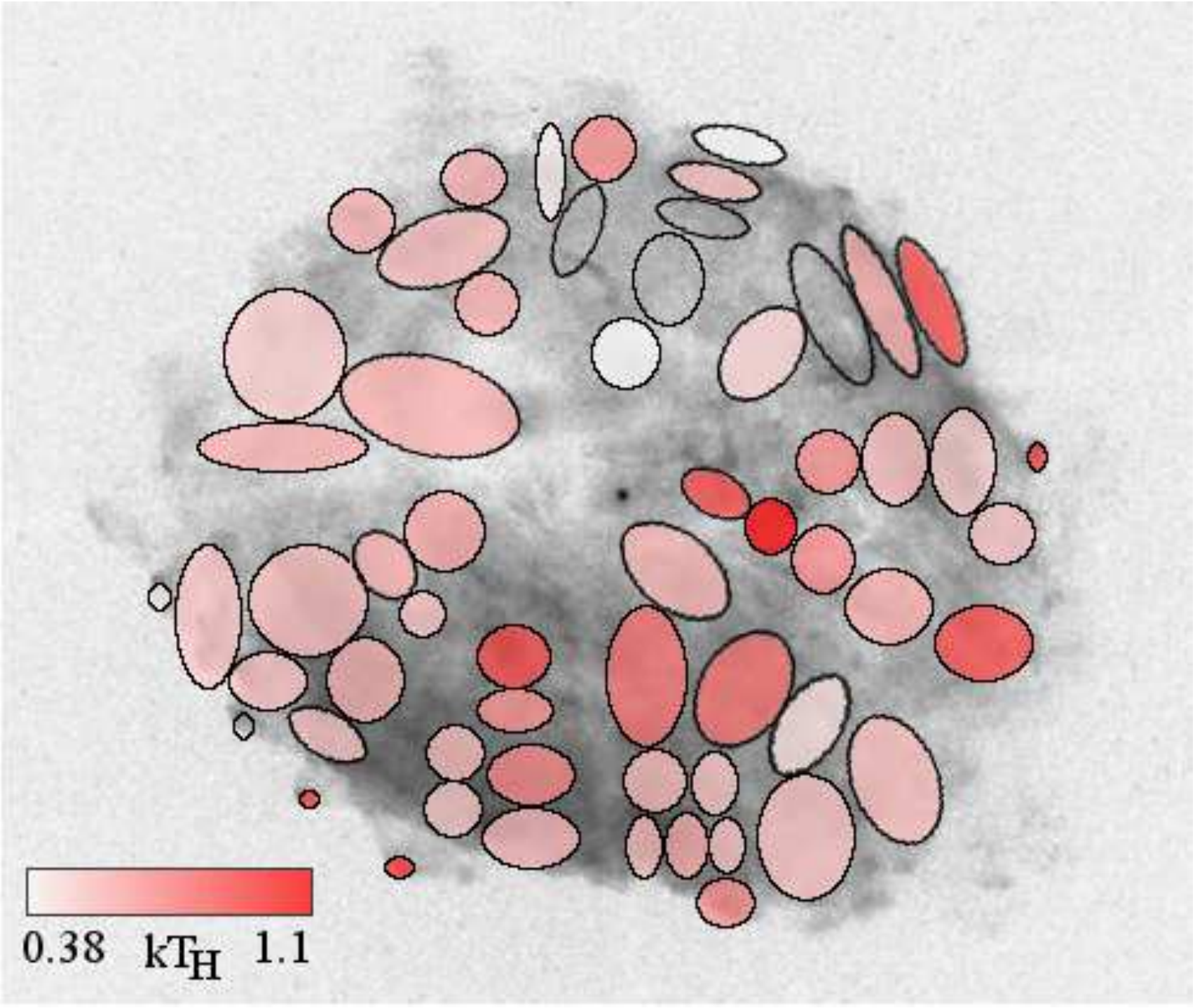}}
		\\
		\subfloat[(d) Temperature (keV; soft)]{\includegraphics[width=0.32\textwidth]{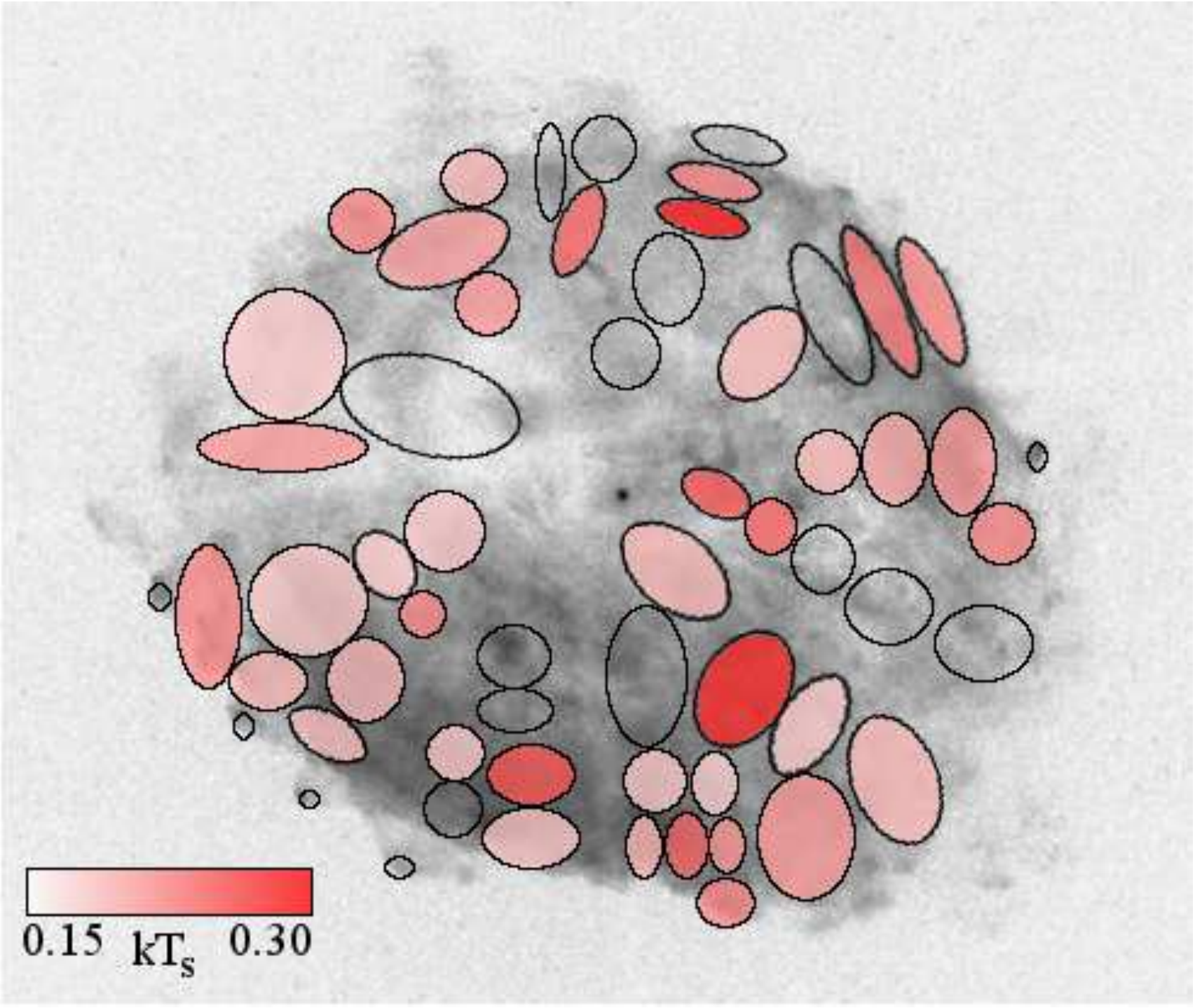}}
		\subfloat[(e) Mg]{\includegraphics[width=0.32\textwidth]{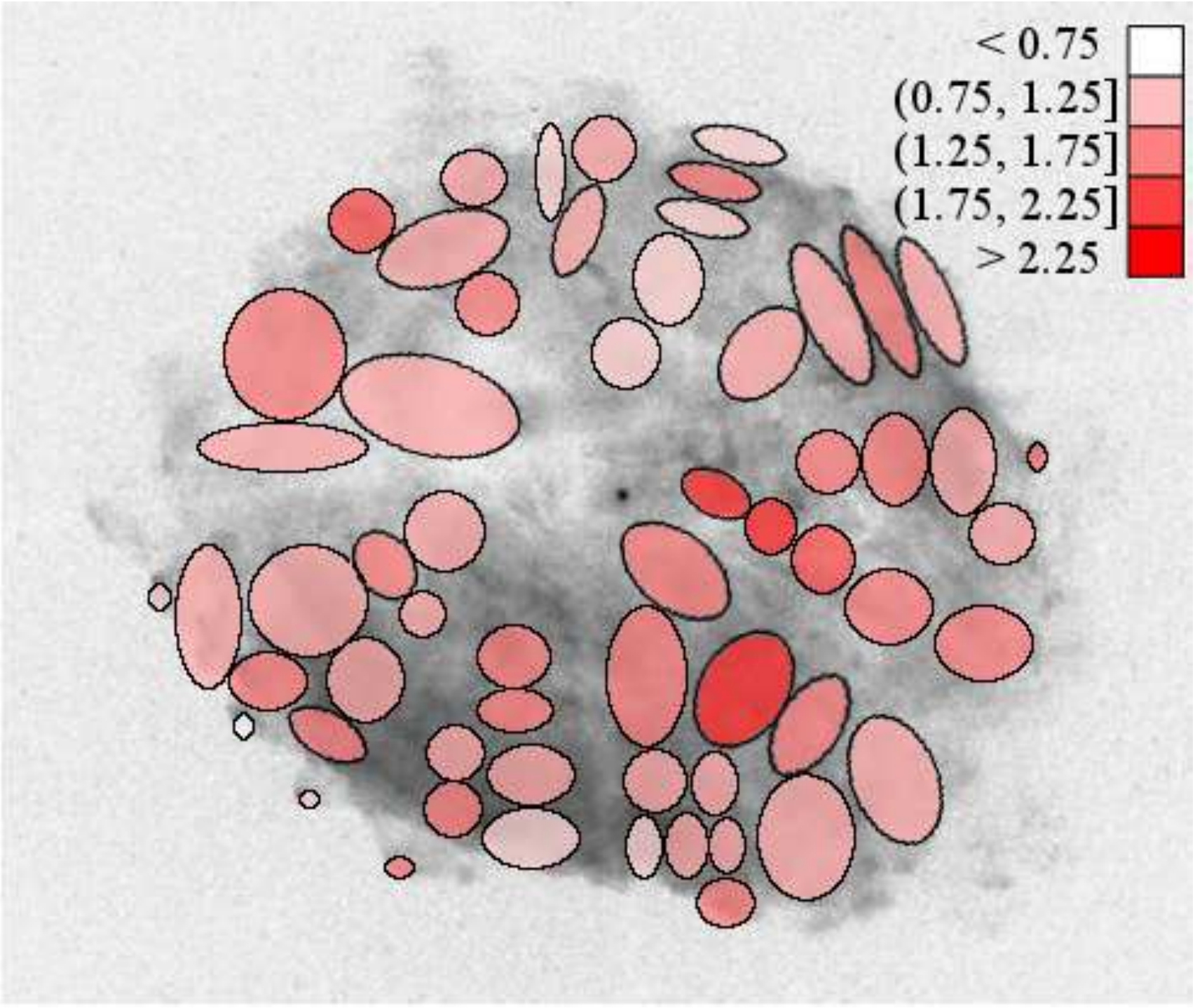}}
		\subfloat[(f) Si]{\includegraphics[width=0.32\textwidth]{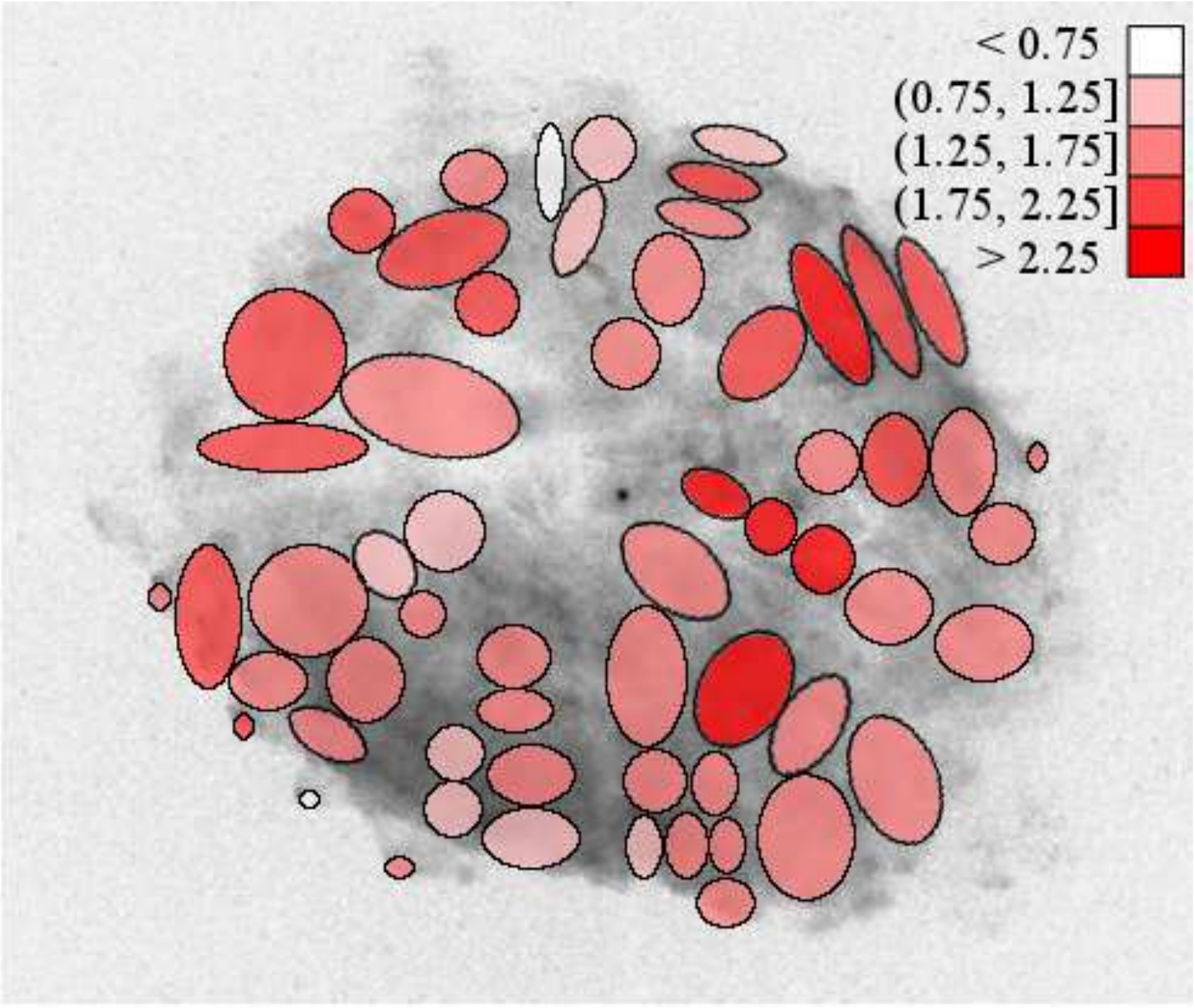}}
	
		\subfloat[(g) S]{\includegraphics[width=0.32\textwidth]{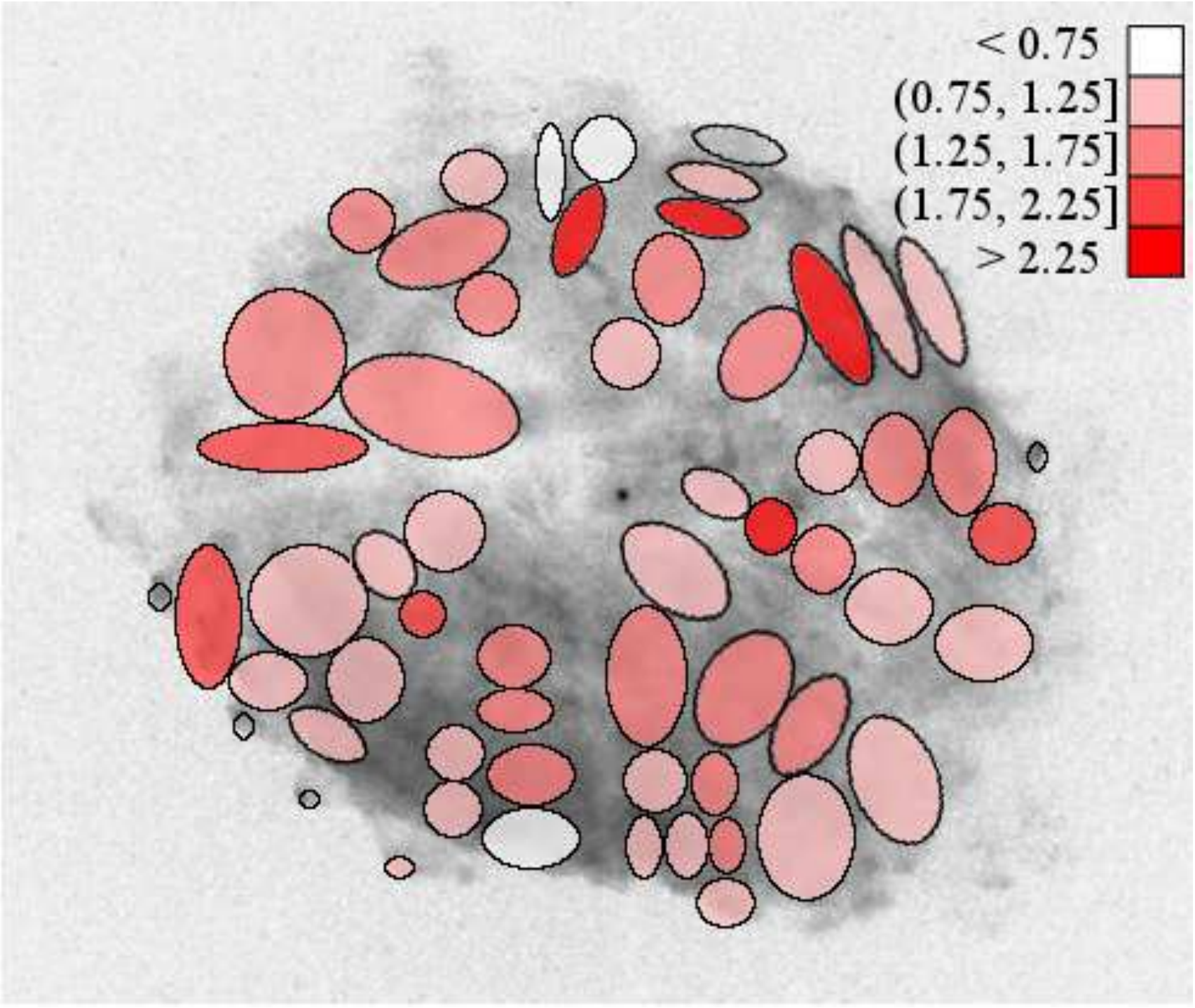}}
		\subfloat[(h) Fe]{\includegraphics[width=0.32\textwidth]{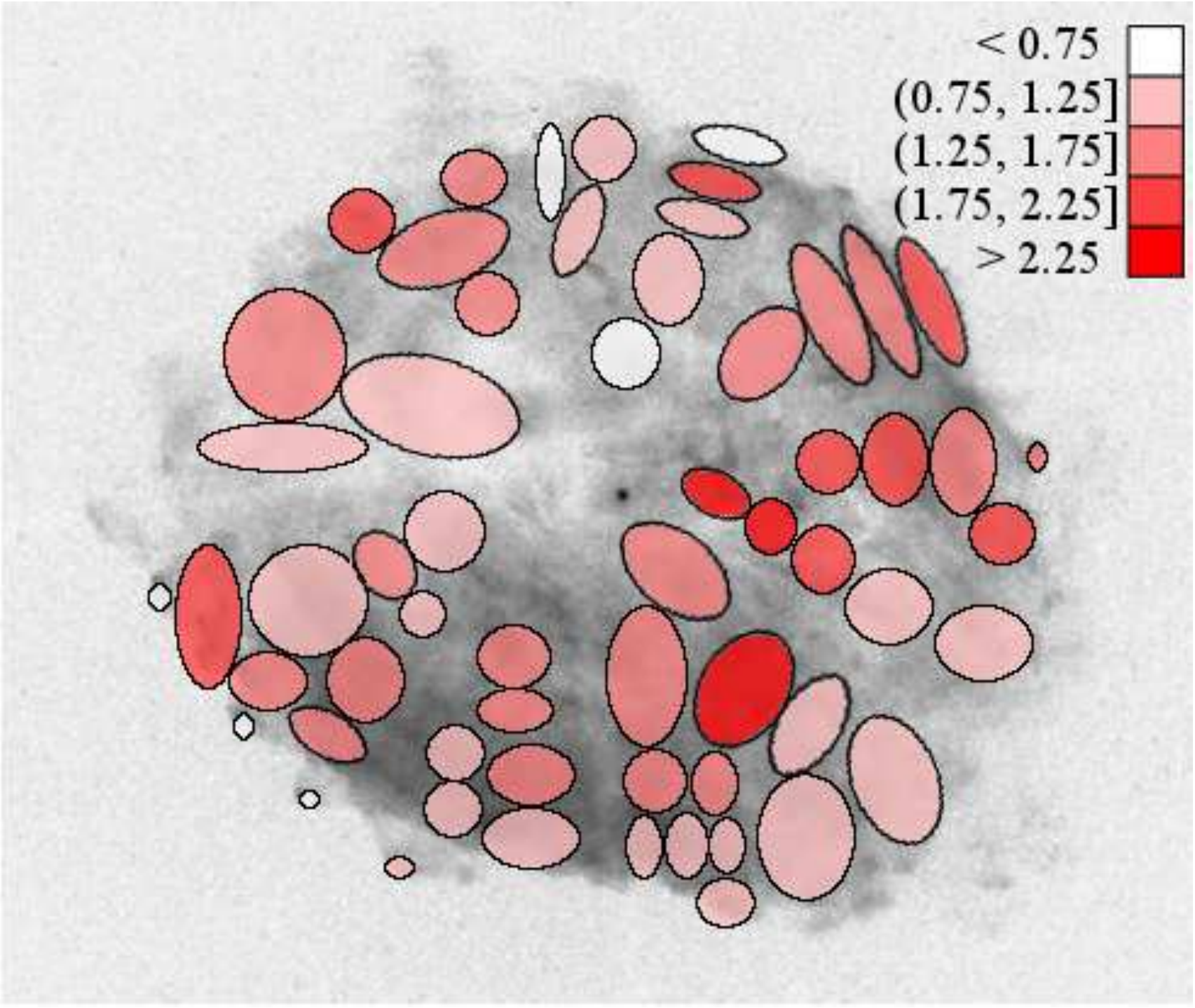}}
		\caption{Parameter maps for each fitted parameter found from Tables~\ref{tbl:OneCompData} and \ref{tbl:RegionData}. Error ranges were not taken into account when creating these images. }
		\label{fig:RegionMaps}
		\end{center}		
	\end{figure*}
	
    \subsection{\textit{XMM-Newton}}

\begin{figure}
		\includegraphics[width=\columnwidth]{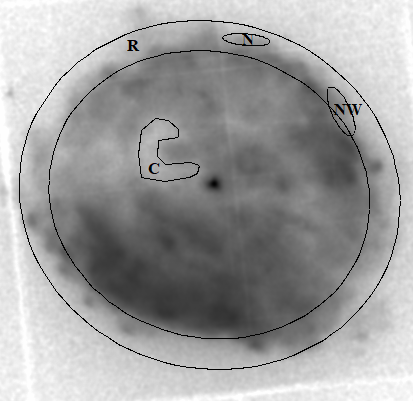}	
		\caption{\textit{XMM-Newton} combined MOS1/MOS2 image from all 3 data sets smoothed with a Gaussian of radius 2 pixels.  Selected low-surface brightness regions are labeled. See \S4.2 and Table~5 for details.}
		\label{fig:xmm}
	\end{figure}

    In this section we used the \textit{XMM-Newton} data to examine the selected low-surface brightness regions of the SNR. The spectra were extracted as described in $\S$2.2 and were fit using the X-ray spectral fitting software XSPEC. Similarly to \textit{Chandra}, the spectral data were restricted to between 0.5--5.0~keV (the ring region was restricted to below 2.0~keV due to poor statistics, especially with data set 701 which was removed for the fitting) and the abundance tables were set using the XSPEC command \textit{abund wilm} \citep{Wilms}. Regions were fit with VPSHOCK and VPSHOCK+APEC models. The region labels and corresponding spectral fits are provided in Fig.~\ref{fig:xmm} and Table~\ref{tbl:xmmSpectra}. There is overall agreement between the \textit{Chandra} and \textit{XMM-Newton} data and the \textit{XMM-Newton} data seem to follow the trends as outlined in the $\S$4.1.3. The ring region encompassing the outer edge of the SNR labelled ``R'' was fit with a single VPSHOCK and is generally consistent with the soft blast wave component obtained from the \textit{Chandra} fits. However, an additional hotter component with enhanced Si improves the fit, suggesting mixing with ejecta. This is evident in the image (Fig.~\ref{fig:xmm}) showing that the ring has a combination of diffuse and knotty/bullet-like emission protruding into the remnant's boundary. The two northern rim regions, the northern most one labelled as ``N'' and the north-western one ``NW'', have the same regions extracted as the \textit{Chandra} regions 8 and 16 respectively. Both regions have spectral properties that are consistent within error with the \textit{Chandra} results. The region labelled ``C'' for the `C-shaped' hole was best fit with a single VPSHOCK consistent with slightly enhanced abundances, and with a relatively high temperature and column density, as expected given its low-surface brightness. This supports its interpretation as an ejecta component with X-ray emission suppressed due to the presence of absorbing material in the foreground. 
   
	\begin{table*} 
    	\caption{Spectra of selected low-surface brightness regions using \textit{XMM-Newton}. For the two-component models, the hard component was fit with the VPSHOCK model and variable abundances, while the soft component was fit with the APEC model and solar abundances. The single-component regions were fit with a VPSHOCK model. }
    
    	\label{tbl:xmmSpectra} 
    	\begin{tabular}{cccccccccc}

    	\hline
    	Region & N$_\text{H}$ & kT & Mg & Si & S & Fe (Ni) & n$_\text{e}$t & kT & $\chi^2_\nu$ (DOF) \\  		  
        & $\times 10^{22}$~cm$^{-2}$ & keV & & & & & $\times 10^{11}$~cm$^{-3}$~s & keV & \\ 	
    	\hline
        R & 1.06$_{-0.03}^{+0.02}$ & 0.30$\pm{0.01}$ & 1.01$_{-0.04}^{+0.05}$ & 2.3$_{-0.2}^{+0.3}$ & ... & 0.75$\pm{0.04}$ & 12$_{-3}^{+5}$ & ... & 1.46 (1346) \\
        N & 1.0$\pm{0.1}$ & 0.45$_{-0.11}^{+0.07}$ & 0.6$\pm{0.1}$ & 0.8$_{-0.3}^{+0.4}$ & ... & 0.46$_{-0.1}^{+0.2}$ & 3.1$_{-1.5}^{+3.7}$ & ... & 1.02 (271) \\
        NW & 1.14$_{-0.05}^{+0.07}$ & 0.6$_{-0.2}^{+0.4}$ & 0.7$\pm{0.7}$ & 1.8$_{-0.6}^{+2.2}$ & 2$_{-1}^{+3}$ & 1.0$_{-0.6}^{+1.3}$ & 2.2$_{-0.7}^{+10.2}$ & 0.22$_{-0.03}^{+0.01}$ & 1.08 (484) \\
        C & 1.16$\pm{0.06}$ & 0.84$\pm{0.08}$ & 1.14$_{-0.1}^{+0.2}$ & 1.6$\pm{0.2}$ & 1.3$_{-0.3}^{+0.4}$ & 0.7$\pm{0.2}$ & 0.9$_{-0.2}^{+0.3}$ & ... & 1.32 (795) \\
    	\hline 	
    	\end{tabular}
    
	\end{table*}
 
\section{Discussion}
	
	We have performed a spatially resolved spectroscopic study of the diffuse emission from SNR RCW~103 using \textit{Chandra} data. The emission from most of the regions can be described by two-component thermal models (VPSHOCK+APEC), with some smaller, more diffuse regions that were fit by one-component thermal models (VPSHOCK or VAPEC). We have experimented with other models made available in XSPEC (such as VNEI) and they provided similar global parameters. In the following section we present a dedicated study of the SNR aimed at addressing the supernova explosion properties. 
	
	\subsection{Blast Wave and Evidence of Ejecta}
	
	The X-ray emission from young SNRs is generally characterized by a forward shock or blast wave component propagating outward into the surrounding ISM and the reverse shock running back into the ejecta. The spatially resolved spectroscopy confirms the necessity of a two-component thermal model to describe the X-ray emission of most regions. The hard component, with plasma temperatures $\sim0.6$~keV, shows slightly enhanced abundance usually associated with the reverse shocked ejecta. The soft component, with plasma temperatures $\sim0.2$~keV, shows abundances at solar usually associated with the blast wave and shocked ISM/CSM. However, some of the one-component regions that were classified as soft had supersolar
	abundances, whereas some hard components had solar or subsolar abundances. Specifically, regions 5, 6, and 8 and their locations at the edge of the shock (see Fig.~\ref{fig:Regions}) imply these are shocked ISM/CSM regions expanding into a lower density medium. This suggests there is a range of temperatures for the shocked ISM/CSM, which makes distinguishing the blast wave and ejecta components challenging. A clear conclusion then is there is indeed mixing between the reverse-shocked ejecta and the shocked ISM/CSM. The ionization timescales for the hard component range from $10^{11}$~cm$^{-3}$~s to $10^{12}$~cm$^{-3}$~s which suggests that the reverse shock is approaching ionization equilibrium (n$_\text{e}\text{t} < 10^{12}$~cm$^{-3}$~s). The high ionization timescale accompanied with the fact that the ejecta yields are relatively low ($\sim 2$) implies that the remnant may be older than previously thought. The soft component for all regions seems to be in CEI. The hard component has slightly enhanced abundance yields and is generally uniform across the remnant as seen from the parameter maps in Fig.~\ref{fig:RegionMaps}. The one-component regions are the only regions with subsolar yields ($< 0.75$), which implies they are shocked ISM/CSM regardless of whether they were classified as hard or soft, whereas the two-component models are consistently about solar or supersolar. 
	
	\subsection{Distance Calculation and Proper Motion}
	
	The distance can be estimated using the column density from the global fit, which is N$_\text{H} = 1.05 \times 10^{22}$~cm$^{-2}$, and a range from the region fits of $(0.74$--$1.40)\times 10^{22}$~cm$^{-2}$. The extinction per unit distance in the direction of RCW~103 is estimated from the colour excess per kiloparsec contour diagrams by \cite{Lucke}. From the diagram, E$_{B-V}/D\sim0.4$~mag~kpc$^{-1}$ and we can calculate the distance relation N$_\text{H}/\text{E}_{B-V}=5.55\times10^{21}$~cm$^{-2}$~mag$^{-1}$ as described by \cite{Schmitt}. We derive a distance of 4.7~kpc, and a range of 3.3--6.3~kpc. The best current distance estimate comes from a radio velocity study of H1 absorption derived a value of 3.1~kpc \citep{Reynoso}. Our estimates are on the high end, which may be due to the higher N$_\text{H}$ values inferred from using \textit{abund wilm} in XSPEC. Thus we subsequently scale our calculations to the parameter D$_{3.1}=D/3.1$~kpc.
	
	We conducted a proper motion study between \textit{Chandra} data sets ObsID~123 and ObsID~17460 taken at two different epochs which span $\sim$16~years from 1999 to 2015 (see Table~\ref{tbl:ExpTime}). A difference image was created to find the best location for motion within the shock similar to Kepler's SNR study by \cite{Katsuda}. The two images were aligned by matching sources detected by the \textit{wavdetect} command from the \textit{CIAO} software as described in the \textit{CIAO} Correcting Absolute Astrometry science thread. From there, the images were normalized across the detectors using the \textit{fluximage} command with a bin of 0.984$^{\prime\prime}$, then normalized based on exposure time, then subtracted. The difference image did not show a clear expansion signature between the two epochs. Additionally, lowering the bin size by a factor of two reduced the image quality prohibiting us from an accurate proper motion measurement. If we consider an expansion rate of 1100~km~s$^{-1}$ inferred from the optical study (\cite{Carter}) and a distance of 3.1~kpc, then we should see an approximate shift of 1.2$^{\prime\prime}$. This should be detectable but was not observed. We then constructed radial profiles separated by 0.5$^{\prime\prime}$ scale on our binned 0.984$^{\prime\prime}$ image for 2 different regions placed perpendicular to the shock front, one at the northern brightened limb and one at the southern limb (see procedure from \cite{Katsuda}). No discernible shifts in position were detected. However, based on the binning of the image, we place an upper limit on the expansion rate of the shock as 900~km~s$^{-1}$, which is comparable to the optical speed. Future epoch \textit{Chandra} observations should allow a better constraint or a measurement of the proper motion.
	
	\subsection{X-Ray Properties of RCW~103}
	
	In the following we derive the X-ray properties of RCW~103 seen in Table~\ref{tbl:XrayProp} assuming first that the shocked ejecta component is described by the hard component and the blast wave component is described by the soft component. In our calculations we assume a radius of 10$^{\prime}$ as estimated by the extent of the radio contours seen in Fig.~\ref{fig:Radio} and a distance of 3.1~kpc which will be written in terms of the scaling factor introduced previously as D$_{3.1}=D/3.1$~kpc. The corresponding physical size is R$_\text{s} = 4.5$~D$_{3.1}$~pc $=1.4\times10^{19}$~D$_{3.1}$~cm. We used the temperatures and derived densities from the full SNR fits in our calculations as they are good representative average of the results from the regions from Fig.~\ref{fig:Regions}. The average and standard deviation of the soft n$_\text{e}$ and n$_0$ values from the small-scale regions of Fig.~\ref{fig:Regions} are $8\pm{3}$~$f_s^{-1/2}$~D$_{3.1}^{-1/2}$~cm$^{-3}$ and $1.6\pm{0.7}$~$f_s^{-1/2}$~D$_{3.1}^{-1/2}$~cm$^{-3}$ respectively, which is in agreement with the values calculated for the full SNR region (see Table~\ref{tbl:XrayProp}). Furthermore, from the Sedov global fits we derived an n$_\text{e}=3.2$~$f_s^{-1/2}$~D$_{3.1}^{-1/2}$~cm$^{-3}$ and n$_0 = 0.67$~$f_s^{-1/2}$~D$_{3.1}^{-1/2}$~cm$^{-3}$, which gives us a shock age of T$_\text{shock}=\tau/\text{n}_\text{e}=2800$~$f_s^{1/2}$~D$_{3.1}^{1/2}$. 
	
	The X-ray emitting regions from Fig.~\ref{fig:Regions} have a volume \textit{V}, estimated by assuming the plasma fills an ellipsoid with semi-major and semi-minor axes from the extracted regions and with a depth along the line of sight equal to the radius of the SNR. The emission measure (EM) is the amount of plasma available to produce the observed flux and is given by $EM = \int n_en_HdV \sim fn_en_HV$ where $n_e$ is the post-shock electron density, and $n_H$ is the proton density ($n_e=1.2n_H$ assuming cosmic abundances). A volume filling factor (\textit{f}) is introduced, which indicates the fraction of emitting plasma that fills the volume of the selected regions. For the separate hard and soft components we consider two distinct filling factors, \textit{f$_h$} and \textit{f$_s$}, respectively. For strong shocks following the Rankine-Hugoniot jump conditions and under the assumption of cosmic abundances, we can estimate the ambient density $n_0$ from the electron density such that $n_e=4.8n_0$; here $n_0$ includes only hydrogen \citep{Borkowski2,Samar&Borkowski}. The normalization factor, $K=\frac{10^{-14}}{4\pi D^2}\int n_en_HdV$, is a parameter from the spectral fits and can be used to calculate the EM. The derived X-ray parameters are summarized in Table~\ref{tbl:XrayProp}. The total absorbed flux of the entire SNR in the 0.5--5.0~keV energy range is $1.8\times10^{-10}$~erg~cm$^{-2}$~s$^{-1}$, the unabsorbed flux is $3.8\times10^{-9}$~erg~cm$^{-2}$~s$^{-1}$, and the total luminosity is  $4.4\times10^{36}$~D$_{3.1}^2$~erg~cm$^{-2}$~s$^{-1}$.
	
	\subsubsection{Uniform Ambient Medium}
	
	The evolutionary phase of the SNR is important for determining the age. By assuming the extreme case of the SNR being in the early free expansion phase of its evolution, we can infer a lower estimate for its age. With an initial expansion velocity of $v_0\sim5000$~km~s$^{-1}$ consistent for young SNRs (see \cite{Reynoso}) and given the size of the remnant, we determine a lower age limit of 880~D$_{3.1}$~yr. When the swept-up mass ($M_{\text{sw}}$) becomes comparable to the ejected mass (M$_{\text{ej}}$), the remnant enters the Sedov phase. Assuming a uniform ambient medium (with the caveat that we should expect some dependency on the radius, as discussed further below and in \S5.3.2), we estimate the swept-up mass from the global SNR fit as $M_{\text{sw}}=1.4m_pn_0\times(4/3\pi $R$^3_sf)=16$~$f_s^{-1/2}$~D$^{5/2}_{3.1}$~M$_\odot$. The age of a remnant is given by t$_{\text{SNR}}=\eta R_s/V_s$, where $\eta=0.4$ for a shock in the Sedov phase \citep{Sedov}. Using the upper limit on the speed of the shock as determined from the proper motion study (\S5.2), we determine an upper age limit of t$_{\text{SNR}}=2.0$~$D_{3.1}$~kyr. However, it is also possible to derive the shock speed from the shock temperature, $T_s$, such that $V_s=(16k_BT_s/3\mu m_h)^{1/2}$ where $\mu=0.604$ is the mean mass per free particle of a fully ionized plasma and $k_B=1.38\times10^{-16}$~erg~K$^{-1}$ is the Boltzmann's constant. Normally, the model component with abundances at solar values is considered the blast wave component. However, in the case of RCW~103 the distinction between the blast wave and ejecta components is not obvious and a range of temperatures may be associated with the blast wave (given the range of ambient densities). With this in mind we calculate the blast wave velocity using the soft component's temperature leading to an upper age limit of 4.4~D$_{3.1}$~kyr. This is consistent with the age estimates obtained from studies at other wavelengths between 2000--4000~yrs (\cite{Mcdonnell}; \cite{Paron}; \cite{Andersen}; \cite{Oliva1999}; \cite{Frail}).   

	The Sedov blast wave model in which a supernova with explosion energy, E$_*$, expands into the surrounding ISM of uniform density $n_0$ is given by E$_*=(1/2.02)R^5_sm_nn_0t^{-2}_{\text{SNR}}$, where $m_n=1.4m_p$ is the mean mass of the nuclei, $m_p$ is the mass of a proton, and $t_\text{SNR}$ is the age of the remnant. The calculated explosion energy using the derived Sedov age from the soft component's blast wave velocity is E$_*=3.7\times10^{49}$~$f_s^{-1/2}$~D$^{5/2}_{3.1}$~erg. We obtained similar results from the \textit{XMM-Newton} ring region fit, with an age of 3.3~D$_{3.1}$~kyr and explosion energy of $2.8\times10^{49}$~$f_s^{-1/2}$~D$^{5/2}_{3.1}$~erg, and from the Sedov global fit with an age of 3.2~D$_{3.1}$~kyr and explosion energy of $4.4\times10^{49}$~$f_s^{-1/2}$~D$^{5/2}_{3.1}$~erg. This is much lower than the canonical value of $10^{51}$~erg but is likely partly due to the under-estimated shock velocity. As well, recent studies of young SNRs where the plasma is generally not in ionization equilibrium show that the electron temperature can be much smaller than the ion temperature (see \cite{Ghavamian}). The modelled temperature parameter from our spectral fits is the electron temperature, and the general assumption that it is equal to the ion temperature may not, in fact, hold. This would lead to an underestimate of the shock velocity and subsequently an underestimate of the explosion energy. It should be noted, however, that lower explosion energies ($\sim10^{50}$~erg) are 
	also found for other SNRs (see e.g., \cite{2012ApJ...754...96K}; \cite{2019MNRAS.482.1031G}).
	In some instances when the shock is expanding into a lower density medium, the hard component, like those described in Section 5.1, can be due to the blast wave component. If we consider the hard, northern regions with temperature of 0.5~keV as representative of the blast wave, 
	we infer a shock velocity of $V_S=670$~km~s$^{-1}$, an age of t$_\text{SNR}=2600$~yr, and an explosion energy of E$_*=5.6\times10^{49}$~$f_h^{-1/2}$~D$^{5/2}_{3.1}$~erg. 
	
    \subsubsection{SNR Expanding in a Wind-Blown Bubble}
	
	We subsequently consider the model of the SNR expanding into the wind bubble of its massive progenitor, a late red-supergiant (RSG), which follows an $r^{-2}$ profile as described by \cite{Chevalier} and considered for other SNRs (\cite{Castro}; \cite{2014ApJ...781...41K}). We assume the SNR is in the adiabatic phase when inferring its expansion velocity and age. \cite{Cox} analytically solved the Sedov profiles for a blast wave expanding in a $r^{-2}$ wind density distribution. \cite{Chevalier} describes the circumstellar wind density as $\rho_{cs}=\dot{M}/4\pi r^2 v_w=Dr^{-2}$ with a dimensionless parameter $D_*=D/D_{ch}$, where $D_{ch}=1\times 10^{14}$~g~cm$^{-1}$ is the coefficient of the density profile, $\dot{M}$ is the mass loss, and $v_w$ is the RSG wind velocity. The mass swept-up by the SNR shock to a radius \textit{R} is given by M$_{\text{sw}}=9.8D_*(R/\text{ 5 pc})$\,M$_\odot$, with the SNR blast wave expanding to a radius $R=(3E/2\pi D)^{1/3}t^{2/3}$. In this adiabatic approximation, the corresponding velocity is given by $V_s=2R/3t$ (\cite{Chevalier1982}; \cite{Chevalier}; \cite{Castro}). For the total SNR emission measure $EM \approx 4.2\times 10^{59}$~$f^{-1/2}_s$~D$^{5/2}_{3.1}$~erg, where we assume the soft component corresponds to the shocked circumstellar medium that fills the full volume of $V=1.1\times 10^{58}$~cm$^3$, we obtain a swept-up mass of M$_{\text{sw}}\approx61$~$f^{-1/2}_s$~D$^{5/2}_{3.1}$~M$_\odot$. This allows for an estimate of the coefficient of the density profile $D=6.3\times10^{14}$~D$^{3/2}_{3.1}$~g~cm$^{-1}$. From \cite{Cox}, the ratio between the average temperature and the shock temperature (weighted by $n^2$) in the uniform case is $\langle T\rangle/T_s=1.276$, and in the RSG wind scenario is $\langle T\rangle/T_s=5/7$. Under the assumption that the X-ray spectrum mostly depends on the average temperature, we can determine a relation between the uniform and $r^{-2}$ density profiles as $T_{\text{s,wind}}/T_{\text{s,uniform}}=1.786$ and $V_{\text{s,wind}}/V_{\text{s,uniform}}=\sqrt{1.786}$. Now the expansion speed, $V_s=\eta R/t$, where for the Sedov case, $\eta=2/3$, and for the adiabatic wind phase, $\eta=2/5$, leading to an estimate for the age ratio of $t_{\text{s,wind}}/t_{\text{s,uniform}}=(5/3)/\sqrt{1.786}$. This leads to a shock speed of $V_{\text{s,wind}}=530$~km~s$^{-1}$ and SNR age of $t_{\text{s,wind}}=5.5$~D$_{3.1}$~kyr. We can then infer the explosion energy as	E$=1.2\times10^{50}$~$f^{-1/2}_s$~D$^{5/2}_{3.1}$~erg. Again, considering the hard component as the blast wave component, we get a wind bubble explosion energy of E$_*=1.7\times10^{50}$~$f_h^{-1/2}$~D$^{5/2}_{3.1}$~erg.

    We conclude that the explosion energy inferred from our X-ray spectroscopy is low ($\leq~2\times10^{50}$~$f^{-1/2}$~D$^{5/2}_{3.1}$~erg) in comparison to standard explosion energies assumed for supernovae, regardless of the assumptions made on the evolutionary stage, ambient environment and exact blast wave temperature.
    
	\begin{table*}
        \caption{Derived x-ray properties of RCW~103. The subscripts ``s'' and ``h'' refer to the soft and hard components, respectively. $EM$: Emission measure, n$_\text{e}$: electron density, $n_0$: ambient density, and $f$: filling factor. Error is to $2\sigma$, with the exception of the full SNR data which has no error.} 
		
			\label{tbl:XrayProp}
		\begin{tabular}{cccccc}	
			\hline
			Region & $EM_h$ & ${\text{n}_\text{e}}_h$ & $EM_s$ & ${\text{n}_\text{e}}_s$ & ${n_0}_s$ \\
			 	   & $\times 10^{56}$~$f_h$~D$_{3.1}^{-2}$~cm$^{-3}$  
			 	   & $f_h^{-1/2}$~D$_{3.1}^{-1/2}$~cm$^{-3}$
			 	   & $\times 10^{57}$~$f_s$~D$_{3.1}^{-2}$~cm$^{-3}$ 
			 	   & $f_s^{-1/2}$~D$_{3.1}^{-1/2}$~cm$^{-3}$
			 	   & $f_s^{-1/2}$~D$_{3.1}^{-1/2}$~cm$^{-3}$ \\
			\hline			
            1   & $1.0\pm{0.3}$             & $1.7_{-0.3}^{+0.2}$       & $0.4^{+1.5}_{-0.2}$       & $3.2^{+6.2}_{-0.7}$   & $0.7^{+1.3}_{-0.2}$   \\
            2   & $1.6^{+0.9}_{-0.1}$       & $2.1^{+0.7}_{-0.6}$       & $2.3^{+1.2}_{-1.5}$       & $8\pm{2}$             & $1.7^{+0.4}_{-0.5}$   \\
            3   & $4.0^{+4.6}_{-1.5}$       & $2.0^{+1.2}_{-0.4}$       & $7.7^{+1.9}_{-3.3}$       & $9^{+1}_{-2}$         & $1.9^{+0.2}_{-0.4}$   \\
            4   & $0.9^{+0.8}_{-0.5}$       & $1.6^{+0.7}_{-0.4}$       & $1.1^{+2.5}_{-0.7}$       & $5^{+6}_{-2}$         & $1.1^{+1.2}_{-0.4}$   \\ 
            5   & $5.2^{+6.8}_{-0.6}$       & $4.6^{+3.0}_{-0.3}$       & ...                       & ...                   & ...                   \\
            6   & $1.1^{+0.5}_{-0.3}$       & $1.6^{+0.3}_{-0.2}$       & ...                       & ...                   & ...                   \\
            7   & ...                       & ...                       & $2.5^{+1.3}_{-1.1}$       & $7\pm{2}$             & $1.6^{+0.4}_{-0.3}$   \\
            8   & $3.5^{+3.9}_{-1.7}$       & $3.0^{+1.7}_{-0.7}$       & ...                       & ...                   & ...                   \\
            9   & $3.1^{+1.7}_{-1.4}$       & $2.8^{+0.8}_{-0.6}$       & $2.5^{+3.3}_{-1.1}$       & $8.0^{+5.4}_{-1.8}$   & $1.7^{+1.1}_{0.4}$    \\
            10  & ...                       & ...                       & $1.4^{+0.6}_{-0.3}$       & $6.1^{+1.2}_{-0.6}$   & $1.3^{+0.3}_{-0.1}$   \\
            11  & $0.39^{+0.26}_{-0.16}$    & $0.8^{+0.3}_{-0.2}$       & $1.8^{+0.3}_{-0.2}$       & $5.4^{+0.4}_{-0.3}$   & $1.12\pm{0.07}$      \\    
            12  & $8.3^{+2.8}_{-2.6}$       & $4.1\pm{0.7}$            & ...                       & ...                   & ...                   \\
            13  & $2.8^{+2.4}_{-1.4}$       & $1.9^{+0.8}_{-0.5}$       & $5.2^{+6.6}_{-2.5}$       & $8^{+5}_{-2}$         & $1.7^{+1.1}_{-0.4}$   \\
            14  & $54^{+2}_{-1}$                     &  $8_{-1}^{+2}$   & ...       & ...         & ...  \\
            15  & $3.3^{+0.7}_{-0.1}$       & $2.2^{+0.7}_{-0.3}$       & $4.6^{+1.5}_{-1.2}$       & $8\pm{1}$             & $1.7^{+0.3}_{-0.2}$   \\
            16  & $0.39^{+0.41}_{-0.09}$    & $0.8^{+0.5}_{-0.3}$       & $1.3^{+0.2}_{-0.4}$       & $3.9^{+2.3}_{-0.6}$   & $0.8^{+0.5}_{-0.1}$   \\
            17  & $6.5^{+3.2}_{-2.0}$       & $1.9^{+0.5}_{-0.3}$       & $0.9^{+2.1}_{-0.3}$       & $2^{+13}_{-2}$        & $0.4^{+2.6}_{-0.4}$   \\   
            18  & $4.3^{+10.2}_{-1.1}$      & $1.6^{+1.0}_{-0.2}$       & ...                       & ...                   & ...                   \\
            19  & $3.4^{+1.1}_{-1.3}$       & $2.0^{+0.3}_{-0.4}$       & $2.1^{+2.4}_{-1.0}$       & $5^{+3}_{-1}$         & $1.0^{+0.6}_{-0.2}$   \\   
            20  & $1.0^{+0.5}_{-0.8}$       & $1.5^{+0.4}_{-0.6}$       & $1.4^{+2.8}_{-0.9}$       & $6^{+6}_{-2}$         & $1.2^{+1.2}_{-0.4}$   \\ 
            21a & $5.3^{+0.9}_{-1.2}$       & $2.9^{+0.2}_{-0.3}$       & $5.2^{+1.9}_{-1.6}$       & $9^{+2}_{-1}$         & $1.9\pm{0.3}$         \\
            21b & $10.6^{+1.4}_{-1.5}$      & $3.7^{+0.2}_{-0.3}$       & $5.8^{+1.8}_{-1.1}$       & $8.8^{+1.4}_{-0.8}$   & $1.8^{+0.3}_{-0.2}$   \\
            22  & $2.5^{+1.1}_{-1.3}$       & $2.5^{+0.5}_{-0.6}$       & $1.3^{+2.0}_{-0.7}$       & $5^{+4}_{-2}$         & $1.1^{+0.9}_{-0.3}$   \\
            23  & $2.5^{+1.1}_{-1.3}$       & $1.6\pm{0.4}$             & $2.4^{+1.7}_{-0.6}$       & $5.1^{+1.8}_{-0.6}$   & $1.1^{+0.4}_{-0.1}$   \\ 
            24  & $13^{+23}_{-12}$          & $3^{+3}_{-1}$             & $11^{+30}_{-10}$          & $9^{+12}_{-4}$        & $1.9^{+2.5}_{-0.8}$   \\
            25a & $1.0^{+0.1}_{-0.7}$       & $1.5^{+0.9}_{-0.5}$       & $1.6^{+9.4}_{-1.0}$       & $6^{+18}_{-2}$        & $1.3^{+3.7}_{-0.4}$   \\  
            25b & $1.8^{+0.1}_{-0.7}$       & $2.6^{+0.9}_{-0.5}$       & $1.8^{+0.7}_{-0.7}$       & $8.3^{+1.6}_{-1.5}$   & $1.7^{+0.4}_{-0.3}$   \\
            26  & $4.5^{+1.4}_{-0.8}$       & $2.6^{+0.4}_{-0.2}$       & $1.8^{+4.6}_{-0.9}$       & $5^{+7}_{-1}$         & $1.1^{+1.4}_{-0.3}$   \\
            27  & $3.2^{+0.7}_{-0.5}$       & $2.6^{+0.3}_{-0.2}$       & $2.8^{+1.6}_{-1.0}$       & $8^{+2}_{-1}$         & $1.6^{+0.5}_{-0.3}$   \\
            28a & $11.8^{+5.6}_{-1.6}$      & $4.3^{+0.2}_{-0.3}$       & $10.7^{+3.1}_{-6.5}$      & $13^{+1}_{-2}$        & $2.6^{+0.2}_{-0.4}$   \\
            28b & $5.0^{+0.4}_{-0.6}$       & $3.6^{+0.4}_{-0.2}$       & $0.5^{+0.4}_{-0.2}$       & $3.7^{+0.6}_{-0.5}$   & $0.8^{+0.3}_{-0.1}$   \\
            29a & $1.0^{+0.1}_{-0.7}$       & $1.6^{+0.9}_{-0.6}$       & $1.5^{+9.4}_{-1.0}$       & $6\pm{2}$             & $1.3^{+0.4}_{-0.2}$   \\  
            29b & $3.3_{-0.6}^{+3.3}$       & $3.1^{+1.6}_{-0.3}$       & $0.7^{+0.5}_{-0.4}$       & $4.6\pm{0.5}$         & $1.0\pm{0.1}$         \\
            30a & $7.9^{+0.6}_{-1.2}$       & $4.6^{+0.2}_{-0.3}$       & $4.1^{+2.4}_{-1.1}$       & $11^{+3}_{-1}$        & $2.2^{+0.6}_{-0.3}$   \\
            30b & $6.7_{-0.8}^{+0.5}$       & $4.3_{-0.3}^{+0.1}$       & $10\pm{8}$                & $17\pm{7}$            & $3\pm{1}$            \\
            33a & $5.7^{+0.9}_{-1.4}$       & $3.2^{+0.2}_{-0.4}$       & $6.0^{+3.6}_{-0.3}$       & $10.3^{+1.3}_{-0.9}$  & $2.1^{+0.3}_{-0.2}$   \\
            33b & $13.4^{+1.1}_{-0.9}$      & $4.6\pm{0.2}$             & $7.7^{+1.9}_{-2.0}$       & $11\pm{1}$            & $2.3\pm{0.3}$         \\   
            34  & $7.4_{-6.8}^{+1.8}$       & $2.8_{-1.3}^{+0.3}$       & $10.4_{-0.3}^{+0.4}$      & $10.6\pm{2}$          & $2.2_{-0.4}^{+0.5}$   \\
            35  & $5.6_{-1.0}^{+1.7}$       & $2.2_{-0.2}^{+0.3}$       & $5.1_{-1.3}^{+1.0}$       & $6.5_{-0.8}^{+0.7}$   & $1.4^{+0.1}_{-0.2}$   \\
            36a & $2.3_{-0.7}^{+1.1}$       & $1.4_{-0.2}^{+0.3}$       & $5.1_{-0.9}^{+1.0}$       & $6.6^{+0.6}_{-0.5}$   & $1.4\pm{0.1}$         \\
            36b & $11_{-2}^{+11}$           & $3.8_{-0.4}^{+1.9}$       & $7.6_{-2.5}^{+12.7}$      & $10_{-2}^{+8}$        & $2.1_{-0.4}^{+1.8}$   \\
            37a & $8.1_{-1.4}^{+1.3}$       & $4.1^{+0.3}_{-0.4}$       & $12\pm{3}$                & $16\pm{2}$            & $3.3_{-0.4}^{+0.5}$   \\
            37b & $2.9_{-0.9}^{+2.0}$       & $3.2_{-0.5}^{+1.1}$       & $6.4_{-1.1}^{+14}$        & $15^{+2}_{-1}$        & $3.1\pm{0.3}$         \\
            38a & $8.0_{-1.2}^{+1.0}$       & $5.8\pm{0.4}$             & $4.4_{-1.8}^{+2.1}$       & $13\pm{3}$            & $2.8^{+0.7}_{-0.6}$   \\
            38b & $5.6_{-2.2}^{+1.7}$       & $4.4_{-0.9}^{+0.7}$       & $1.9_{-0.8}^{+1.0}$       & $8\pm{2}$             & $1.7\pm{0.4}$         \\
            38c & $4.1\pm{1.0}$             & $4.4\pm{0.5}$             & $1.3_{-0.8}^{+1.8}$       & $7.8_{-2.3}^{+5.2}$   & $1.6_{-0.6}^{+1.1}$   \\
            39  & $2.0_{-0.9}^{+1.1}$       & $2.6^{+0.7}_{-0.6}$       & $1.8_{-0.8}^{+1.6}$       & $8_{-2}^{+4}$         & $1.6_{-0.4}^{+0.7}$   \\
            40  & $12_{-1}^{+2}$            & $3.1\pm{0.2}$             & $6.6_{-1.7}^{+2.8}$       & $7.1^{+1.5}_{-0.9}$   & $1.5_{-0.2}^{+0.3}$   \\
            41  & $5.9_{-0.9}^{+1.1}$       & $2.2\pm{0.2}$             & $2.7_{-1.2}^{+1.9}$       & $5_{-1}^{+2}$         & $1.0_{-0.2}^{+0.4}$   \\
            42  & $0.7_{-0.4}^{+0.6}$       & $1.4_{-0.5}^{+0.6}$       & $1.3_{-0.7}^{+1.1}$       & $6_{-2}^{+3}$         & $1.3_{-0.4}^{+0.5}$   \\
            43  & $0.22_{-0.18}^{+0.80}$    & $0.9_{-0.3}^{+1.6}$       & $1.5_{-0.7}^{+0.9}$       & $7\pm{2}$             & $1.5\pm{0.4}$         \\
            44  & $1.7_{-0.3}^{+0.6}$       & $0.6_{-0.6}^{+1.1}$       & ...                       & ...                   & ...                   \\
            45  & $5.5_{-1.5}^{+1.6}$       & $2.8\pm{0.4}$             & ...                       & ...                   & ...                   \\
            46  & $5.4_{-1.2}^{+1.1}$       & $2.6\pm{0.3}$            & ...                       & ...                   & ...                   \\
            Bullet 1 & $1.1_{-0.7}^{+2.2}$      & $3_{-1}^{+3}$         & ... & ... & ...  \\
            Bullet 2 & ...   & ...  & $0.013_{-0.003}^{+0.004}$ & $1.1_{-0.1}^{+0.2}$ & $0.22\pm{0.01}$  \\
            Bullet 3 & $0.14_{-0.07}^{+0.14}$   & $0.9_{-0.2}^{+0.4}$   & ... & ... & ...  \\
            Bullet 4 & $0.4_{-0.2}^{+0.6}$      & $2.2_{-0.5}^{+1.5}$   & ... & ... & ...  \\
            Bullet 5 & $1.34_{-0.05}^{+2.81}$   & $3.8_{-0.6}^{+3.9}$   & ... & ... & ...  \\
            SNR & 518.6 & 2.0 & 416.2 & 5.7 & 1.2  \\       
			\hline	
		\end{tabular}	
	\end{table*}
	
\subsection{Progenitor Mass}

    The nucleosynthetic yields can be used constrain both the progenitor mass and the supernova explosion energy.  Here we compare the nucleosynthetic yields from a range of progenitor plus explosion models to the abundances of the ejecta component obtained from the fitted X-ray data found in Table \ref{tbl:RegionData}. The abundance ratios shown are ratios with respects to Si given by ($X$/Si)/($X$/Si)$_{\odot}$, where $X$ is the measured ejecta mass of either Mg, S, or Fe with respect to Si and with respect to their solar values from \cite{Wilms}.  Here we use 4 sets of models for comparison: bipolar explosion models \citep{M03} hereafter labelled as M03, hypernova models \citep{N06} hereafter labelled as N06, a suite of spherical explosions using a range of progenitor masses \citep{S16} hereafter labelled S16, and a recent set of explosion models using 3 progenitors but with a broad range of explosions \citep{fryer18}, hereafter labelled F18.  These models span a range of progenitor masses and supernova explosion properties.

\begin{figure*}
		\begin{center}
			\includegraphics[width=0.6\textwidth]{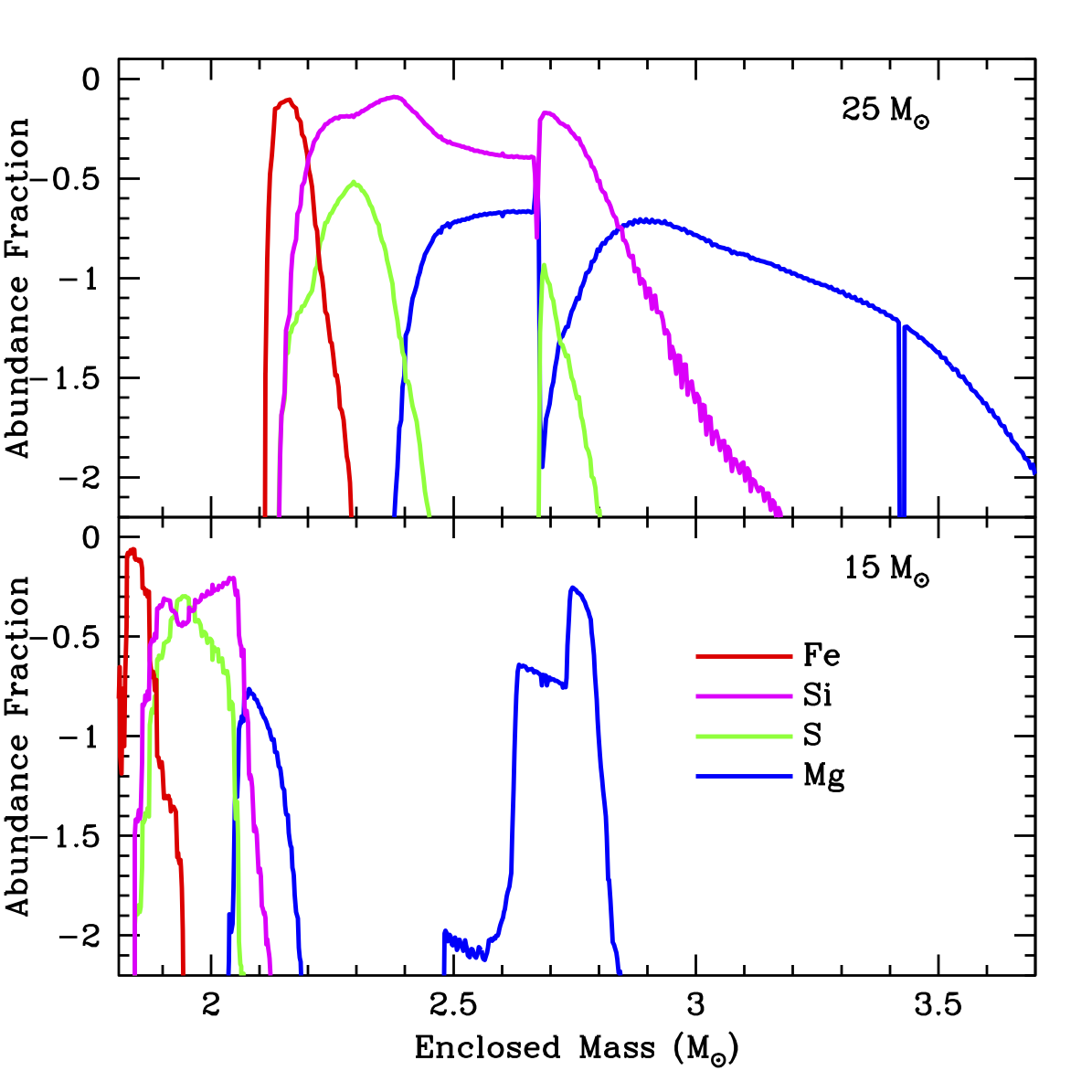}	
			\caption{Abundance fraction for magnesium, silicon, sulfur and iron as a function of enclosed mass for a 15 and 25\,M$_\odot$ stars under fairly typical (near $10^{51} {\rm \, erg}$) explosion energies \citep{fryer18}.  Iron is produced in the innermost ejecta when the supernova shock fuses material in the silicon layer.  It both depends sensitively on the shock strength and the fallback and is an excellent tracer of the explosion energy.  The other elements are both destroyed and created when the supernova shock passes through the star and are better tracers of the progenitor star.}
			\label{fig:abun}
		\end{center}
	\end{figure*}

    To understand how the yields constrain the progenitors and supernova engines, let's review the sites of these yields.  Fig.~\ref{fig:abun} shows the distribution of 4 key elements (Mg, S, Si, Fe) as a function of enclosed mass for two different supernova explosions (15,25\,M$_\odot$ progenitors).  The iron (decay product from $^{56}$Ni) is produced when the supernova shock is driven through the silicon layer.  Both because it is produced by the shock and because it is produced in the innermost ejecta and hence is sensitive to the fallback, the iron production is very dependent upon the explosion energy.  Produced just above the region producing iron, sulfur also depends sensitively on the shock strength, but is less likely sensitive to fallback.  The other elements are produced in the stellar burning shells with further production/destruction when the supernova shock passes through the burning shells.  These elements are more dependent upon the stellar structure.  In this manner, we can constrain both the progenitor mass and explosion energy.

\begin{figure*}
		\begin{center}
    		\subfloat[(a) Maeda et al. (M03) Model]{\includegraphics[width=0.5\textwidth]{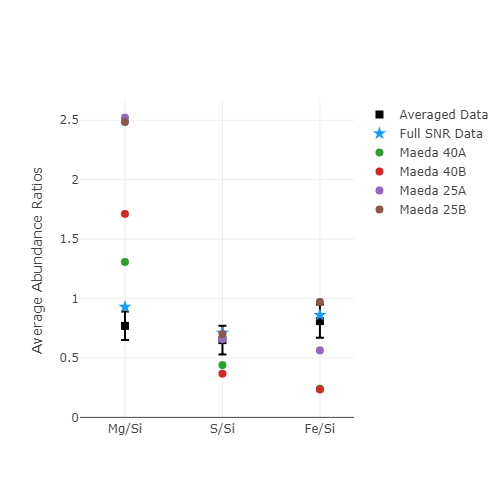}}
    		\subfloat[(b) Nomoto et al. (N06)  Model]{\includegraphics[width=0.5\textwidth]{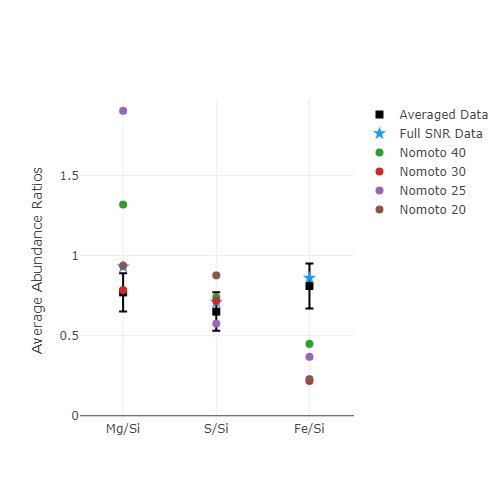}}
    		\\
            \subfloat[(c) Sukhbold et al. (S16) Model]{\includegraphics[width=0.5\textwidth]{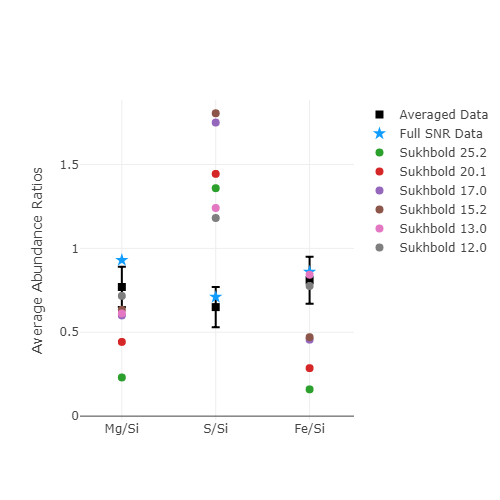}}
    
    		\caption{Best-fit abundances for Mg, S, and Fe relative to Si relative to solar values for the averaged abundance values from Table \ref{tbl:RegionData} using a root mean-square for the error bars (black square) and for the full SNR fit (blue star). Three core-collapse nucleosynthesis models with predicted relative abundances [$X$/Si]/[$X$/Si]$_{\odot}$ are over-plotted for the models M03, N06, and S16 with different masses labeled and in units of $M_{\odot}$. }
    		\label{fig:Progenitor}	
		\end{center}
	\end{figure*}

    Fig.~\ref{fig:Progenitor} shows a series of comparisons between the model yields (N06, M03, and S16 model suites) and the averaged yields from the observations.  The observed abundances from all the regions in Table \ref{tbl:RegionData}, except for the full SNR fits, were averaged and used a root mean square for the error. We did not report errors for the full SNR fit (given the large reduced $\chi^2$), but show the global fit values (labelled as the blue star in the figure) for comparison. The model results span a range of stellar masses.  The size of the silicon layer (and hence the mass of the silicon ejecta), increases with progenitor mass.  The amount of iron ejecta does not increase in a commensurate manner.  These models struggle to produce the high Fe/Si ratio in RCW~103.  The best fit models tend to be the lower-mass progenitors, e.g. the lowest mass progenitors from S16.
 
\begin{figure*}
		\begin{center}
			\includegraphics[width=0.6\textwidth]{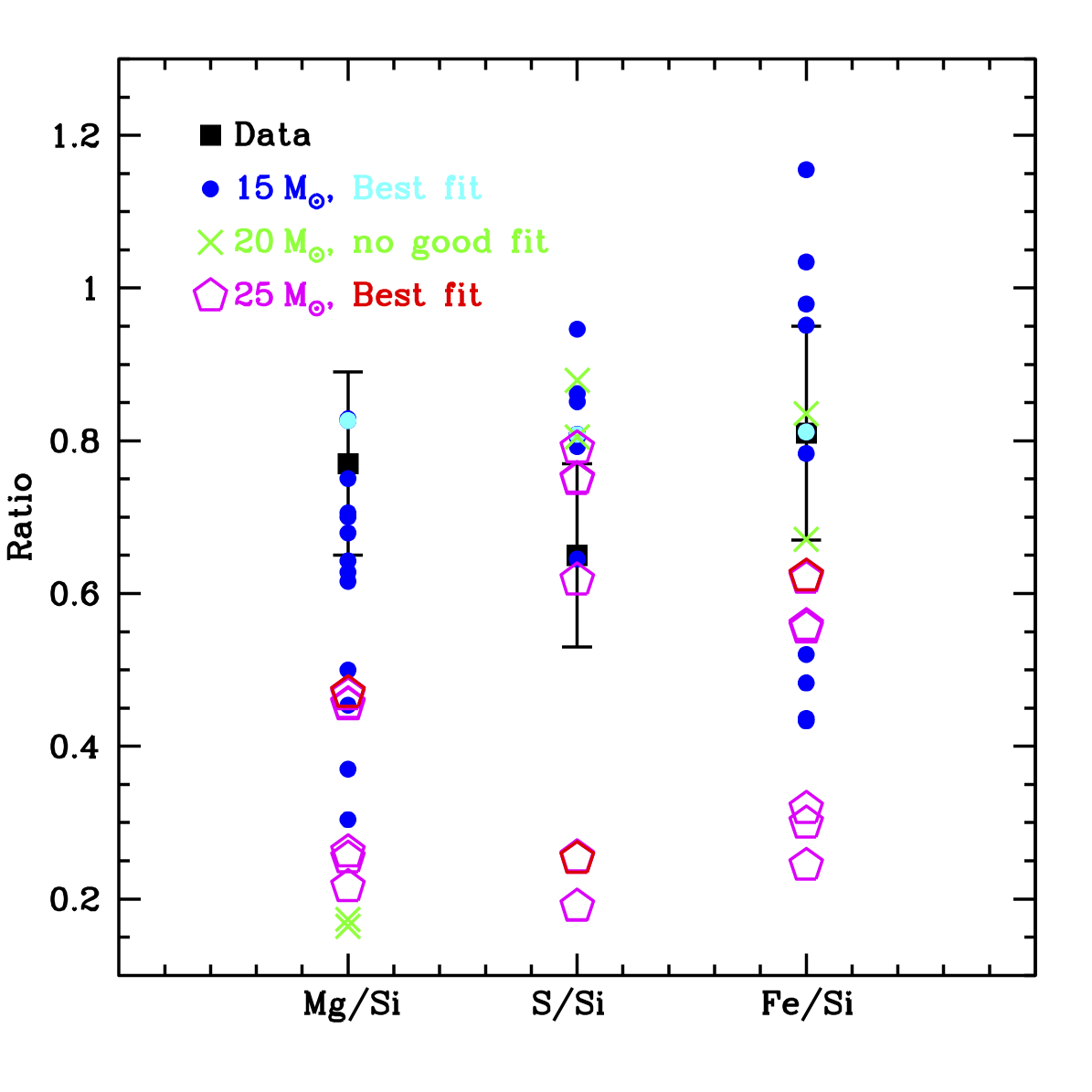}	
			\caption{Like Figure \ref{fig:Progenitor}, the best-fit abundances for Mg, S, and Fe relative to Si for the F18 models that vary the explosion energy.  The blue circles correspond to 15\,M$_\odot$ progenitor explosions (the cyan is the best fit statistically), the crosses correspond to 20\,M$_\odot$ progenitor explosions (no good fits) and the magenta pentagons correspond to the 25\,M$_\odot$ progenitor explosions (the red is the best fit). In agreement with the S16 models, the lowest-mass F18 models (in this case, a 15\,M$_\odot$ progenitor) fits the data better.  There exist explosion models with the 15\,M$_\odot$ progenitor that are good fits to the data.  The best fit 25\,M$_\odot$ explosion produces too little sulfur and iron.  Although some 25\,M$_\odot$ explosions can produce enough sulfur, none produce enough iron. }
			\label{fig:SNexpyield}
		\end{center}
	\end{figure*}

\begin{figure*}
		\begin{center}
			\includegraphics[width=0.6\textwidth]{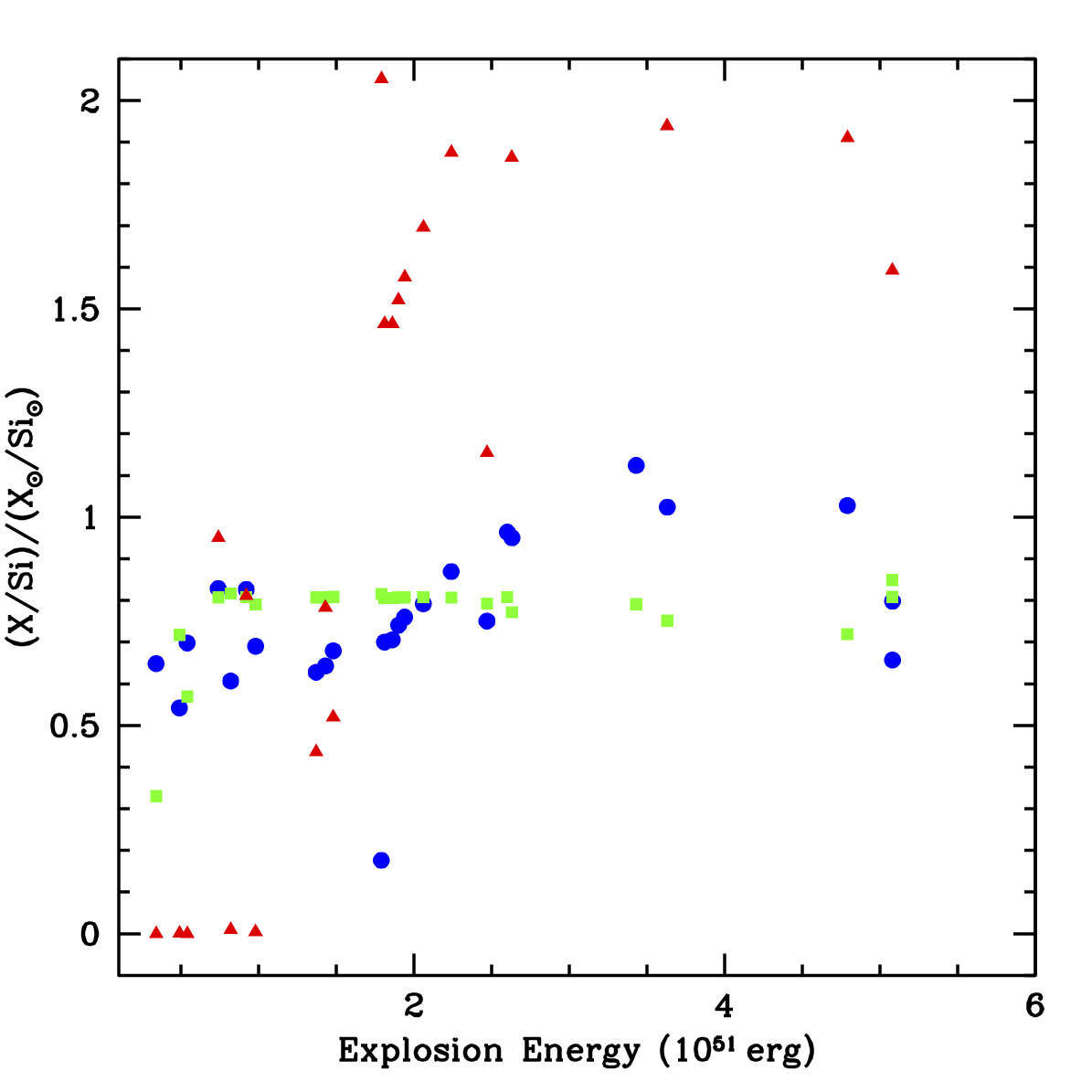}	
			\caption{Yield ratios for the F18 15\,M$_\odot$ models as a function of explosion energies.  The best fits to the data have explosions in the energy range from $0.5-2.0\times10^{51} {\rm \, erg}$.  In these 1-dimensional explosion models, it is difficult to eject much $^{56}$Ni if the explosion energy is much below $0.5\times10^{51} {\rm \, erg}$. The colours of the datapoints are the same as those shown in Fig.\ref{fig:abun}.}
			\label{fig:yve}
		\end{center}
	\end{figure*}

	These models use a fixed explosion energy for each progenitor mass.  The F18 database has a broad set of explosion energies for its 3 progenitors.  Figure \ref{fig:SNexpyield} shows the fits to the data with these models.  With this range of energies, fits can be found for progenitor masses that did not fit well from the earlier fixed-energy models (e.g. S16).  Good fits for all elements are found with a 15\,M$_\odot$ progenitor.  To get a better handle on the fits to the data, we show the yields for all of the F18 15\,M$_\odot$ progenitor models (Figure \ref{fig:yve}).  The models that best match the data have explosion energies lying in the range of $0.5-2.0\times10^{51} {\rm \, erg}$.  In these 1-dimensional models, it is difficult to eject much $^{56}$Ni if the explosion energy is below this value.  Even with extensive mixing from strongly asymmetric explosions \citep{hungerford03}, it will be difficult to explain the iron yield for explosions with energies below $0.5\times10^{51} {\rm \, erg}$.  
	
	All of these models assume progenitors at solar metallicity.  Metallicity can alter the yields of the stars as well, changing the size of the silicon layer, but these changes are typically small compared to the sensitivity to the explosion energy and differences between different progenitor models and masses.  For example, the silicon shell in the \cite{woosley02} models changes by 0.1\,M$_\odot$ from a solar metallicity model to a $10^{-4}$ solar metallicity model, increasing the total silicon mass by nearly 15\%.  It is difficult to say for certain how this will alter the yield ratios, but it is likely that similar explosion energies will produce similar iron yields for both models.  Hence, the iron to silicon ratio will decrease by 15\%.
	
	\subsection{Comparison to Previous Studies}
	
	A previous \textit{Chandra} study on RCW~103 has been reported by \cite{Frank}. In the following we compare our studies in detail. Firstly, Frank et al. focused primarily on the progenitor's mass using only the \cite{N06} progenitor model which they determined to be 18--20\,M$_\odot$. This paper is a detailed study using additional unpublished archived data aimed at determining all intrinsic properties of the engine driving this remnant, namely the explosion energy, ambient density, age, expansion velocity, and distance. Furthermore, for the supernova progenitor science, we have used the latest suite of improved/new core collapse nucleosynthesis models, particularly the \cite{S16} models and the \cite{fryer18} models, when inferring the progenitor's mass. From the Sukhbold et al. 2016 model we determined a progenitor mass of 12--13\,M$_\odot$ in contrast to Frank et al.'s 18--20\,M$_\odot$. There are 4 main points differentiating the methodology used in the two works: 1) the focus of the individual studies, 2) the regions selection and model fits, 3) the background subtraction, and 4) the assumptions made by Frank et al. on the CSM abundances. 
	
	Frank et al. examined 27 regions selected based on their \textit{Chandra} EWI line emission study, whereas in this paper, a total of 54 regions were selected for a more complete coverage of the entire SNR. Frank et al. considered single-component VPSHOCK model fits whereas we found that a two-component model (VPSHOCK+APEC) was statistically required for many regions. Multi-temperature plasma is expected from SNRs due to contributions from the ejecta and CSM/ISM components. This has been shown to be the case for many ejecta-dominated SNRs (e.g. \cite{2014ApJ...781...41K}; \cite{Samar&Borkowski}). We also extended the energy coverage to beyond what was used by Frank et al., who restricted their analysis to the 0.5--3.0~keV range, whereas we considered a broader energy range of 0.5--5.0~keV, which impacted the spectral properties and required a multi-component analysis for most regions. As a result the abundance values have been impacted and our analysis supports evidence of mixing between the shocked ISM/CSM and ejecta.
	
	For the background subtraction, we experimented with different background regions for each region selection, to ensure the backgrounds were of the same size as the region, landed on the same CCD chip that the region is located (regardless of dataset), and relatively nearby to the region. This is important to minimize contamination by the Galactic ridge emission. When this was not possible, mostly for the 970 dataset, a background region was selected from the same chip as the region is located. We have also tested different backgrounds for any variation in the results due to potential contamination. 
	
	Finally, for the progenitor study, the presence of enhanced ejecta is essential for determining the mass of the progenitor. Both studies remarked on the difficulty of separating the ejecta and blast wave components. Because Frank et al. found no enhanced ejecta above solar values nor used two-component models, they argued that specific regions with the lowest abundances are representative of circumstellar regions at roughly 0.5 times the solar value. They subsequently set this as the abundance value for the shocked CSM, and anything above 0.5 times the solar would be considered ejecta. We did not make this assumption, but rather considered above-solar abundances as evidence of ejecta. Furthermore, we used the most recent core-collapse nucleosynthesis models and discussed degeneracies in inferring the progenitor mass due to e.g. different explosion energies.

	Despite some differences in the two studies, our results agree on many fronts. Our equivalent width images obtained with \textit{XMM-Newton} showed overall similar global trends as the \textit{Chandra} images of Frank et al., with a relatively uniform Fe~L, and anti-correlated lobe structure for Mg, Si, and S. Variations across the SNR with respect to column density, N$_\text{H}$, are shown in both studies to be lowest in the south-west and highest in the north-east. The temperature and ionization timescales match up well with the hard component of the two-component fits in our study, with average temperature values of 0.60~keV and ionization timescale of $10^{11}$--$10^{12}$~cm$^{-3}$~s. Frank et al. reported post-shock electron densities with their highest values in the south-west limb, similar to the results in our study. Finally, the regions from Frank et al. found their CSM regions in ionization equilibrium, which parallels the soft component of this study. This strengthens the argument that the soft component of the VPSHOCK+APEC fits from this work primarily describes the blast wave and has reached ionization equilibrium.

    \subsection{Linking the SNR to the CCO}

    Our spatially resolved spectroscopic study of the remnant suggests a low-energy explosion into a low-metallicity environment.  If the explosion energy is below $0.5 \times 10^{51} {\rm \, erg}$, much of the iron would fall back onto the compact remnant, making it impossible to match the observed abundance data.  It is unlikely that large asymmetries can mix out enough iron to explain the iron abundances.  However, if the compact remnant re-ejects this material (perhaps through a magnetized CCO), this material can be ejected, matching both the low explosion energies and the iron yield. 
    
    We note that some of our results on RCW~103 (low progenitor mass and low explosion energy) agree with a most recent study dedicated to a sample of SNRs hosting magnetars \citep{Ping2019}.

\section{Conclusions}
	
	This work presents the first dedicated, complete and deepest study of the SNR RCW~103 through a full imaging and spatially resolved X-ray spectroscopic study performed using 135.9~ks and 79.6~ks of \textit{Chandra} and \textit{XMM-Newton} data, respectively. Our study was aimed at determining the intrinsic properties of the supernova explosion and the physical properties of the remnant hosting the peculiar CCO 1E~161348--5055. The region selections are found in Fig.~\ref{fig:Regions}, with the spectral fits summarized in Table~\ref{tbl:OneCompData} and \ref{tbl:RegionData}, and the derived X-ray properties found in Table~\ref{tbl:XrayProp}. Below is a summary of our study. 
	
	\begin{enumerate}
	    \item The high-resolution X-ray images confirm a spherical morphology with diameter $\sim$10$^{\prime}$, and two brightened limbs in the south-east and north-west. The southern limb is mostly soft in the east and harder in the south-west, with knots of multi-band regions throughout. The bright northern limb is not as soft as its southern counterpart, and contains more medium-energy X-ray emission with a few multi-band regions. The SNR interior has small, clumpy features, with the north-east side more diffuse than the rest of the SNR. A peculiar `C-shaped' hole centrally located just north-east of the hard CCO. The north east regions tend to have the highest column density, N$_\text{H}$, which indicates that the emission from this region, including the hole, is absorbed by some foreground material. The radio contours mimic the X-ray morphology including the brightened limbs and depression north-east of the CCO.
	    \item The line images obtained with the \textit{XMM-Newton} data show overall a similar distribution to the broadband images for the Fe L, Mg, and Si lines, with enhanced emission in the limbs. The equivalent width maps show a more uniform distribution across the SNR, with some depression in some regions; however these maps are limited by low count statistics.
	    \item The \textit{Chandra} emission from the SNR is dominated by thermal X-ray emission from plasma with kT~$<1$~keV, best fitted by a single-component VPSHOCK model, or for most regions by a two-component VPSHOCK+APEC model. The two-component models have a relatively hard ($\text{kT}\sim0.60$~keV) VPSHOCK component, still not yet in ionization equilibrium with ionization timescales n$_\text{e}=10^{11}$--$10^{12}$~cm$^{-3}$~s, with  slightly super-solar abundances. The soft ($\text{kT}\sim0.20$~keV) component had reached CIE, with solar or sub-solar abundances. This suggests that the hard component is dominated by the reverse shock heated ejecta, whereas the soft component is dominated by the forward shock. The single component models were subsequently sorted into hard or soft, based on their temperatures. However, there was some difficulty in separating the ejecta and blast wave components of the regions, which indicates there is likely a wide range of temperatures associated with the blast wave and which is attributed to a non-uniform ambient density. The N$_\text{H}$ values were highest in the north-east and lowest in the south-east limb, whereas the ionization timescales and abundances had no obvious distribution across the remnant. 
	    \item A distance and proper motion study were performed using the \textit{Chandra} data. The column density from the full SNR fit and the extinction contour diagrams by \cite{Lucke} were used to calculate a distance of 4.7~kpc, and a range of 3.3--6.3~kpc. The most recent distance range from \cite{Reynoso} using an HI absorption study found a distance of 3.1~kpc. The proper motion study puts an upper limit on the speed of the shock as 900~km~s$^{-1}$.
	    \item The global SNR spectrum was poorly fit by the VPSHOCK+APEC model with a $\chi^2_\nu\approx12$ which indicates the presence of multi-temperature components. We determined a lower age limit of 880~D$_{3.1}$~yr (free-expansion phase) and an upper age limit of 4.4~$D_{3.1}$~kyr (Sedov phase). The Sedov phase yields a shock velocity of $400$~km~s$^{-1}$ (assuming the soft component represents the blast wave component), leading to a swept-up mass of M$_{\text{sw}}=16$~$f_s^{-1/2}$~D$_{3.1}^{5/2}$~M$_\odot$ and an explosion energy of E$_*=3.7\times10^{49}$~$f_s^{-1/2}$~D$^{5/2}_{3.1}$~erg under the assumption of an explosion in a uniform ambient density. When considering expansion into a stellar wind, the explosion energy is E$_*=1.2\times10^{50}$~$f^{-1/2}_s$~D$^{5/2}_{3.1}$~erg. By making different assumptions on the shock temperature for the blast wave velocity, we similarly obtain a low energy ($<10^{51}$~erg) explosion. We conclude that the explosion energy inferred from our X-ray spectroscopy is $\leq~2\times10^{50}$~$f^{-1/2}$~D$^{5/2}_{3.1}$~erg, low in comparison to standard explosion energies assumed for supernovae, regardless of the assumptions made on the evolutionary stage, ambient environment and exact blast wave temperature.
	    \item Standard explosion models did not match the ejecta yields for RCW~103. Our best estimate yields a progenitor mass around 12--13\,M$_\odot$ from the Sukhbold et al. 2016 model.  However, altering the explosion energy can produce good fits to the data, even with a 15\,M$_\odot$ progenitor. It is likely that a good fit can be found for lower mass progenitors with the right explosion energy. 
	    \item{Although a range of explosion energies of the 15\,M$_\odot$ progenitor can be found to fit the abundance data, the range of explosion energies tend to lie in the 0.5--$2\times 10^{51} {\rm \, erg}$. A magnetized CCO could possibly re-eject fallback material, allowing lower explosion energies to still match the observed abundances.}
	\end{enumerate}
	
	Future observations of RCW~103 with \textit{Chandra} resolution will enable a measurement of the proper motion that will further constrain the dynamics and energetics of the supernova remnant.
	Better fits would be obtained with a suite of models that include a full set of metallicities, progenitor masses and SN explosion energies.  Such a detailed suite has not been produced in the literature.  In addition, our model comparisons to observations of RCW~103 focused on fitting the average abundances to 1-dimensional explosion models.  The convective supernova engine is expected to produce multiple outflows \citep{herant95,fryer02} with strong and weak shocks.  By studying the fully distribution of yields, we should be able to determine the asymmetries in the explosion.  

\section{Acknowledgements}
This research made use of NASA's Astrophysics Data System and HEASARC maintained at NASA's Goddard Space Flight Center. We thank Ping Zhou
 for input on nucleosynthesis models. We acknowledge the support of the Natural Sciences and Engineering Research Council of Canada (NSERC) through the Canada Research Chairs and the NSERC Discovery Grants programs (S.S.H.) and a Canada Graduate Scholarship (C.B.), and the University of Manitoba's GETS program.  The research by C.F. was supported by the Laboratory Directed Research and Development program of Los Alamos National Laboratory under project number 20190021DR and from the U.S. Department of Energy Office of Science and the Office of Advanced Scientific Computing Research via the Scientific Discovery through Advanced Computing (SciDAC4) program and Grant DE-SC0018297.




\bibliographystyle{mnras}







\bsp	
\label{lastpage}
\end{document}